\documentclass[12pt]{article}
\pdfoutput=1
\usepackage[normalem]{ulem}
\usepackage{epsfig}
\usepackage{diagbox}
\usepackage{graphicx}
\usepackage{subcaption}
\usepackage{amsmath}

\usepackage{amssymb}
\usepackage{verbatim}
\usepackage{bm}
\usepackage{dsfont}
\usepackage{cite}
\usepackage{textcomp}
\usepackage{calc}
\usepackage{geometry}
\usepackage{bbm}
\usepackage{xcolor}
\usepackage[utf8]{inputenc}
\usepackage{ulem}
\usepackage{feynmp}
\usepackage{tikz-feynman}
\tikzfeynmanset{compat=1.1.0}
\usepackage[english]{babel}
\usepackage{mathtools}
\usepackage{graphicx}
\usepackage[affil-it]{authblk}
\usepackage{hyperref}
\usepackage{color}
\usepackage{verbatim}
\usepackage{slashed}
\usepackage{cancel}
\usepackage{braket}
\usepackage{multicol}
\usepackage{tcolorbox}
\tcbuselibrary{theorems}
\usepackage{bm}
\usepackage{enumitem}
\usepackage{float}
\usepackage{soul}

\def\micro{{\tt micrOMEGAs}}

\def\calchep{{\tt CalcHEP}}

\usepackage{hyperref}
\hypersetup{colorlinks=true,urlcolor=blue,linkcolor=red,citecolor=green!60!black}

\newcommand{\be}{\begin{equation}}
\newcommand{\ee}{\end{equation}}
\newcommand{\beq}{\begin{equation}}
\newcommand{\eeq}{\end{equation}}
\newcommand{\bea}{\begin{eqnarray}}
\newcommand{\eea}{\end{eqnarray}}

\newcommand{\mgamma}{m_{\gamma'}}
\newcommand{\mphi}{m_\phi}

\newcommand{\Gg}{g_{\phi\gamma\gamma'}}
\newcommand{\gss}{g_{\star s}}
\newcommand{\gs}{g_{\star }}
\newcommand{\TRH}{T_{\rm RH}}
\newcommand{\Tdec}{T_{\rm decay}}

\newcommand{\adec}{a_{\rm decay}}
\newcommand{\rhoc}{\rho_{c,0}}

\hypersetup{linktocpage=true}

\begin{document}

\begin{center}
\vspace*{0.5cm}
{\LARGE{\textbf{Freezing-in the {Pure} Dark Axion Portal}}}\\
\date{}
\vspace*{1.2cm}

{\large{Paola Arias$^{1}$, Basti\'an D\'iaz S\'aez$^{2}$, and Joerg Jaeckel$^{3}$
}}\\[3mm]
\it{$^1$Facultad de Ingenier\'ia, Universidad San Sebasti\'an,\\
campus Ciudad Universitaria, Santiago, Chile} 
\\

\it{$^2$Instituto de Física, Pontificia Universidad Católica de Chile, \\
Avenida Vicuña Mackenna 4860,
Santiago, Chile} \\
\it{$^3$Institut f\"ur theoretische Physik, Universit\"at Heidelberg, \\
Philosophenweg 16, 69120 Heidelberg, Germany} \\
  
  \end{center}

\date{}

\begin{abstract}
A $Z_2$ symmetry under which ``dark'' and ``visible'' fields transform differently can further seclude already dark particles. In a dark sector consisting of an axion-like particle and a dark photon the dominant interaction can then be via a dark photon-axion portal to the ordinary photon. If the axion is lighter than the dark photon it naturally emerges as a viable dark matter candidate. We explore its freeze-in production both in a standard high-reheating temperature and weak coupling as well as a low-reheating temperature and strong coupling regime. Cosmological constraints are taken into account to identify viable regions in parameter space. For the strong freeze-in regime, we highlight the possibility of probing it at B-factories, beam dump and collider experiments. We also briefly discuss a potential misalignment contribution to the dark matter density and delineate under which conditions it can be neglected.

\end{abstract}

\newpage
\tableofcontents
\newpage

\section{Motivation}
A common definition of a hidden sector is that the new, ``dark'' particles residing in it are uncharged under the Standard Model (SM) gauge symmetries. On the renormalizable level, this limits the interactions to three portals~\cite{Okun:1982xi,Holdom:1985ag,Foot:1991kb,Binoth:1996au,Schabinger:2005ei,Patt:2006fw,Ahlers:2008qc,Batell:2009yf,Beacham:2019nyx}. In many discussions this is often complemented by the dimension 5 axion or axion-like particle (ALP) portal (cf., e.g.~\cite{Agrawal:2021dbo,Antel:2023hkf}) \footnote{Taking ALPs to be pseudoscalar generically makes them harder to detect compared to scalars, thereby providing some extra hiddenness.}. In the former cases, the leading interaction is usually some form of ``mixing'' such as photon-dark photon kinetic mixing. The corresponding mixing is taken to be small and therefore the particles are ``hidden''. In the latter case, the only interaction is via a higher-dimensional operator and suppressed by a large energy scale.

For low mass particles such a suppression is often already
enough to make them sufficiently stable and feebly interacting to be a good dark matter candidate.
For larger particle masses, however, stability often requires fairly extreme values of the relevant parameters. For example, an axion-like particle with a mass of $\gtrsim 100\,{\rm MeV}$ interacting with two photons must have a coupling constant suppressed by a value larger than the Planck scale to survive until today. It is therefore very tempting to consider an additional level of suppression by having a non-trivial symmetry in the dark sector.\footnote{For heavy dark matter candidates it is a very common technique to ensure their stability via an additional symmetry, e.g. {matter or} R-parity in SUSY~\cite{Goldberg:1983nd,Ellis:1983ew} see also, e.g.~\cite{Olive:2003iq,Martin:1997ns} for reviews.} 

In the following, we consider exactly such a situation: a model with an ALP and a dark photon, both non-trivially transforming under a $Z_{2}$ symmetry. In this case the main\footnote{The symmetries also allow for a quadratic Higgs portal interaction. However, we take it to be sub-dominant.} interaction is a ``pure'' dark axion portal, $\sim \phi F^{\mu\nu}F'_{\mu\nu}$. Although we consider primarily an effective model, a concrete realization that gives this type of interaction has been given in~\cite{Hook:2023smg}. Previously, some of us have studied this model in the context of very light dark matter~\cite{Arias:2020tzl} finding a rich phenomenology as well as an extra level of protection from constraints. Cosmological effects of this type of model have been considered in~\cite{Gutierrez:2021gol,Hong:2023fcy,DiazSaez:2024dzx,Broadberry:2024yjw}. Astrophysical effects have been taken into consideration in \cite{Kalashev:2018bra}. The reactor, collider and fixed target phenomenology has been studied in~\cite{deNiverville:2018hrc,Deniverville:2020rbv,Jodlowski:2023yne, Jodlowski:2024lab}. More generally, models with a dark axion portal term, $\sim \phi F^{\mu\nu}F'_{\mu\nu}$, but also including $Z_2$ violating terms, have been discussed in~\cite{Jaeckel:2014qea,Kaneta:2017wfh,Gao:2020wer,Kalashev:2018bra,Choi:2018dqr,Deniverville:2020rbv,Hook:2021ous,Domcke:2021yuz,Gutierrez:2021gol}.

In the present paper we now want to exploit the stability of the dark particles {due to the exact $Z_2$ symmetry} to enable them to be dark matter also at larger masses. We focus on freeze-in~\cite{Hall:2009bx,Elahi:2014fsa, Balazs:2022tjl, Jain:2024dtw} production {(freeze-out has already been discussed in~\cite{DiazSaez:2024dzx}).} As our results show, the freeze-in abundance crucially depends upon the cosmological conditions at the time of production. Notably, the transition from an inflaton-dominated Universe to a radiation-dominated one—known as the reheating era—is often assumed to take place at very early times, yielding very high
temperatures. However, this assumption is not guaranteed and scenarios with low reheating temperatures are possible~\cite{Giudice:2000ex, Hannestad:2004px, Allahverdi:2020bys}. 
 Remarkably, such scenarios have been shown to have a dramatic impact on thermal and nonthermal production mechanisms in the context of axion physics \cite{Grin:2007yg, Visinelli:2009kt,Blinov:2019jqc, Carenza:2021ebx,
 Arias:2021rer, Arias:2023wyg}. Therefore, in this work, we will allow the reheating temperature to be as low as $\TRH\gtrsim 5$~MeV and consider high and low temperature freeze-in scenarios.  

The paper is structured as follows. In the next section~\ref{sec:model} we concretize our model setup and choices. Then, in section~\ref{sec:mis} we discuss potential contributions from misalignment and discuss the regions where they can be negligible, thereby preparing our initial state for the freeze-in production.\footnote{In principle it would be possible to consider a combination of both contributions (cf., e.g.~\cite{Feruglio:2024dnc} for a paper considering this in a different model), but for simplicity we will not do this here.}
The processes contributing to the freeze-in are discussed in section~\ref{sec:freeze}. Section~\ref{sec:cosmo} then discusses the cosmological evolution after the initial production and the constraints that result from it. We also show the final results in the form of the regions of parameter space where we find viable freeze-in axion dark matter. Finally, we conclude in section~\ref{sec:conclusions}.

\section{Model and parameter space}\label{sec:model}

We extend the SM with a new gauge boson $A'_\mu$ from a $U(1)_X$ gauge symmetry, plus a singlet scalar $\phi$. We choose that both fields transform in an odd way under a discrete  $Z_2$ symmetry as $(A'_\mu, \phi) \rightarrow -(A'_\mu, \phi)$, such that the Lagrangian is given by\footnote{One could  allow a mixing between the new $U(1)_X$ gauge boson with the hypercharge one if the $Z_2$ is only approximate \cite{Chen:2024jbr}. We do not follow that possibility here.}
\begin{eqnarray}\label{lag1}
 \mathcal{L} \supset  - \frac{1}{4}F'_{\mu\nu}F^{'\mu\nu} - \frac{1}{2}m_{\gamma'}^2A_\mu^{'2} + \frac{1}{2}(\partial_\mu \phi)^2 - \frac{1}{2}m_\phi^2\phi^2 - \lambda_\phi \phi^4 - \lambda_{HS}\phi^2 |H|^2 + \mathcal L_{F,B},
\end{eqnarray}
where $\mathcal L_{F,B}$ refers to a single higher dimensional operator given by
\begin{eqnarray}\label{five_dim}
\mathcal L_F= \frac{\Gg}2 \phi F_{\mu\nu}\tilde F'^{\mu\nu},\quad \mathcal L_B= \frac{\Gg}2 \phi B_{\mu\nu}\tilde F'^{\mu\nu},
\label{eq:portal}
\end{eqnarray}
with $\Gg$\footnote{Despite the important difference between the coupling to photons and hypercharge bosons above the electroweak scale (see also App.~\ref{app:unitarity}) we call both coupling constants $\Gg$. } a  dimensionfull parameter, $F_{\mu\nu}$ the $U(1)_{\rm EM}$ photon field strength, $B_{\mu\nu}$ the $U(1)_{Y}$ field strength, and $F_{\mu\nu}^{'}$ the {dark} photon (DP) field strength. 

A concrete way to realize such an effective model with a suitable $Z_2$ symmetry based on a renormalizable underlying model  has been given in~\cite{Hook:2023smg}, finding a coupling,
\begin{equation}
\label{eq:coup}
    g_{\phi\gamma\gamma'}\sim \frac{g_yg_d}{\pi^2 f},
\end{equation}
where $f$ is the ``axion'' decay constant, $g_y$ is the hypercharge gauge coupling and $g_d$ the coupling of the dark sector U(1).

Typically, ALPs and {DPs} are studied with masses below the GeV energy scale, such that the physics description is adequate with $\mathcal{L}_{F}$ \cite{Kaneta:2016wvf}. However, our scan of the freeze-in (FI) parameter space also includes high temperatures where the consistent operator is $\mathcal{L}_{B}$ \cite{Heeba:2018wtf}, cf. also Appendix~\ref{app:unitarity}. 

The $Z_2$ symmetry ensures that other portals, such as kinetic mixing or the ordinary axion portal, are shut, thus protecting the stability of the lightest dark particle. 

In addition we take the, in principle possible, Higgs portal coupling to be negligible, inspired by the fact that a shift symmetry forbids non-derivative couplings \cite{Arias:2020tzl}.

In this way, the free parameters of the model are 
\begin{eqnarray}
    \{m_\phi, m_{\gamma'}, \Gg\} .
\end{eqnarray}
We consider a scenario in which both states are massive, such that the lightest is the absolute stable to be the DM candidate whereas the other one decays before the present time\footnote{For instance, if we consider $m_\phi = 0$ and $m_{\gamma'}\sim \mathcal{O}$(MeV), we should consider $\Gg$ in the ballpark of $10^{-20}$ GeV$^{-1}$ for having a stable dark photon on cosmological scales to be the DM. That level of coupling is far below the couplings required for freeze-in, which is the main focus of this work on the DM production mechanism.}. In particular, we focus on the hierarchy $m_\phi < m_{\gamma'}$, although the other one is perfectly viable. 

\section{Preliminaries -- Misalignment and the initial state}\label{sec:mis}
Before looking in detail at the thermal scattering processes that contribute to the freeze in production, let us briefly comment on the initial state.

For freeze-in production it is usually assumed that, after inflation, the universe contains a negligible abundance of the desired dark matter particles~\cite{Hall:2009bx}.
However, for bosonic fields this is not automatically guaranteed as there might also be a contribution from the misalignment mechanism~\cite{Preskill:1982cy,Abbott:1982af,Dine:1982ah,Arias:2012az}, i.e. a contribution from a (spatially constant) displacement of the field from its minimum value that survives inflation. This can lead to a potentially sizable contribution to the dark matter density, e.g. for the case of a scalar it is~\cite{Arias:2012az,Alonso-Alvarez:2019ixv}
{\begin{equation}
\label{eq:misalignment}
\frac{\Omega_{\phi}}{\Omega_{\rm DM}} \sim 5\left(\frac{\phi_{1}}{10^{11}\,{\rm GeV}}\right)^{2}\left(\frac{m_{\phi}}{10\,{\rm keV}}\right)^{1/2}\sim 5\left(\frac{\phi_{1}}{10^{9.5}\,{\rm GeV}}\right)^{2}\left(\frac{m_{\phi}}{10\,{\rm GeV}}\right)^{1/2}.
\end{equation}}
Notably, this grows with the mass.

In our case, we have two new dark sector fields: the ALP and the dark photon.
The latter is relatively harmless. The reason is that for vectors, and as long as we do not invoke non-trivial effects such as, e.g. a sizable extra coupling to the Ricci scalar~\cite{Arias:2012az}, dilution during inflation is effective~\cite{Golovnev:2008cf}. After a suitable period of inflation, the contribution is then negligible.

For scalars this is not the case, as is clear from the above formula~\eqref{eq:misalignment} for a scalar with standard couplings.
However, we can invoke that axion-like particles, being pseudo-Goldstone bosons have a typical field value that is constraint by the corresponding symmetry breaking scale $f$. Moreover, validity of the effective field theory also requires the field value not to be too large. Let us take the following naive estimate,
\begin{equation}
    \phi\lesssim f\sim  c \Gg^{-1},
\end{equation}
where $c$ is a combination of couplings in the underlying theory. For the model of~\cite{Hook:2023smg} this is given by $c\sim g_y g_d/(\pi^2)$ (cf.~Eq.~\eqref{eq:coup}). Taking the dark coupling to be of the same order as the hypercharge coupling, this leads to $c\sim 10^{-2}$, which we take as an indicative value.
Using Eq.~\eqref{eq:misalignment} a coupling of
\begin{align}
\Gg\gtrsim 7\times 10^{-12}\,{\rm GeV}^{-1}
    \left(\frac{c}{10^{-2}}\right)\left(\frac{m_{\phi}}{10\,{\rm GeV}}\right)^{1/4},
\end{align}
ensures that the misalignment contribution is sub-dominant.

As we will see in the following this includes a significant part of the interesting parameter space.
However, there are also sizable regions of freeze-in parameter space where this is not guaranteed.
Importantly, however, Eq.~\eqref{eq:misalignment} only applies if the Hubble scale during inflation is higher than the mass, $3H_{\rm inflation}\gtrsim m_{\phi}$.
If the inflation scale is lower, the scalar dilutes with the volume during inflation, and again, the abundance after inflation is negligible. 
Taking reheating to be reasonably effective we have,
\begin{equation}
    T_{\rm RH}\sim {\mathcal{O}(1)}\times 2\frac{\sqrt{H_{\rm inflation}M_{P}}}{g_{\star}^{1/4}},
\end{equation}
where $g_{\star}$ indicates the number of relativistic degrees of freedom {and $M_P=2.4\times 10^{18}$~GeV, the reduced Planck mass.}
Using this we find the following condition ensures a small initial abundance,
\begin{equation}
\label{eq:freezeincondition}
T_{\rm RH}\lesssim 10^{6}\,{\rm GeV}\left(\frac{200}{{\gs}}\right)^{1/4}\left(\frac{m_{\phi}}{10\,{\rm keV}}\right)^{1/2}.
\end{equation}
As we shall see in the following (cf. Fig.~\ref{fig:gvsmphi_TRH}), this condition is fulfilled in a good fraction of our parameter space of interest.

If we do not fulfill this condition, we still may avoid the misalignment contribution. A simple, though admittedly not very pretty option is fine-tuning the initial value of the ALP. This should be possible as long as the required value to make the misalignment contribution subdominant is not smaller than the inflation scale. This requires,
\begin{align}
    \TRH\ll 2\times 10^{14}~{\rm{GeV}}\left(\frac{100}{\gs}\right)^{1/4}\left(\frac{m_{\phi}}{10\,{\rm keV}}\right)^{-1/8}.
\end{align}
This is a rather mild condition, even for relatively high masses $m_{\phi}$. Therefore, in the following, we will simply assume that the misalignment contribution is small. If required, by tuning the initial field value.

Finally we note that an alternative way would be to invoke suitable extra couplings such as, again, e.g., a non trivial coupling to the Ricci-scalar~\cite{Alonso-Alvarez:2019ixv}.\footnote{Another possible example might be a sizable quartic self-interaction. But this may also affect the thermal behavior.} 
 But in these situations the above caveats on non-trivial model features should be kept in mind.

\section{Freeze-in DM production}\label{sec:freeze}
The freeze-in mechanism is typically based on the fact that the couplings between the DM and the visible sector are weak, so the production of the former never reaches thermal equilibrium \cite{Hall:2009bx}.

Regarding the dark-axion portal, the primary mechanism for dark matter production has been misalignment \cite{Kaneta:2017wfh, Arias:2020tzl}. Freeze-in production in the lightweight mass regime and its implications for BBN and $N_{\rm eff}$ were explored in Ref.\cite{Hong:2023fcy}. The production of dark photon DM through freeze-in, by gluon fusion, was analyzed in Refs.\cite{Kaneta:2016wvf, Gutierrez:2021gol}, while production via the dark Primakoff effect was studied in Ref.~\cite{Kaneta:2017wfh}. 

Before we turn to the different production channels, let us start with a few generalities of the calculation. As already mentioned, in the present work, we neglect the Higgs portal coupling, $\lambda_{HS} \approx 0$, and consider the pure axion portal with coupling $\Gg$ as the only relevant coupling between the dark and visible sectors\footnote{There is a certain threshold for the maximum value of $\Gg$ such that above that value, thermal freeze-out should take place via coscattering (aka conversion driven-freeze-out) \cite{DiazSaez:2024dzx}.}. As discussed in the previous section, we take the initial population of $\phi$ and $\gamma'$ to be negligible for temperatures above the reheating temperature. For our choice $m_{\phi} < m_{\gamma'}$, the total relic abundance of ALPs receives a contribution from the direct production of ALPs during the FI but also one from the decay $\gamma'\rightarrow \gamma(Z) + \phi$,
\begin{eqnarray}\label{par_den}
    \Omega_\phi h^2 = \Omega_\phi^{FI} h^2 + \Omega_\phi^{decay} h^2\,.
\end{eqnarray}
Ultimately, the final relic abundance of $\phi$ depends on the final yield $Y_\phi^f$ after DP decay, 
\begin{eqnarray}\label{relic_high_TR}
    \Omega_{\phi} h^2 = \frac{s_{\infty} Y_\phi^f}{\rho_{\rm c,0}}h^2 \,.
\end{eqnarray} 
 Here, $s_\infty = 1.36\times 10^{-37}$ GeV$^3$ is the present entropy density today, $\rho_{c,0} = 3.65\times 10^{-47}$ GeV$^4$ the critical density, and $h = 0.67$. We take the relic abundance as measured by Planck collaboration $\Omega_{c} h^2 = 0.12$ \cite{Planck:2018vyg}. In order to obtain the relic abundance, we make use of \micro\ 6.0.4 \cite{Belanger:2001fz}.

In the following, we detail the relevant Boltzmann equations with their corresponding integrated collision terms for the production of each dark state. 

\subsection{Production of dark states}
The leading production processes for dark states correspond to $2\rightarrow 2$ reactions, with the initial states being SM particles. Following the conventions for identical particles of \micro\ for freeze-in \cite{Belanger:2018ccd}, the integrated Boltzmann equations for the number density of $\phi$ and $\gamma'$ are
\begin{eqnarray}\label{boltzeq1}
 \dot{n}_\phi + 3Hn_\phi &=& 2 C(\gamma\gamma\rightarrow \phi\phi) + \sum_{f} C(f\bar{f}\rightarrow \phi {\gamma'}) + C(W^+W^-\rightarrow \phi\gamma'), \\  
  \dot{n}_{\gamma'} + 3Hn_{\gamma'} &=&  \label{boltzeq2}2 C(\gamma\gamma\rightarrow {\gamma'}{\gamma'}) + \sum_{f} C(f\bar{f}\rightarrow \phi {\gamma'}) + C(W^+W^-\rightarrow \phi\gamma'),
\end{eqnarray}
where $\gamma, W^\pm$ represent the SM photon and weak charged carriers of the SM, $f(\bar{f})$ SM fermion (antifermion). Here $H = \sqrt{\frac{\pi^2g_*}{90}}\frac{T^2}{M_P}$, with $M_P$ the reduced Planck mass of $2\times 10^{18}$ GeV, and on the right side of each equation the integrated collision terms are given by~\cite{Belanger:2018ccd},
\begin{eqnarray}
    C(A\rightarrow B) &=& \int\prod_i \left(\frac{d^3p_i}{(2\pi)^3 2E_i} f_i\right)\prod_j \left(\frac{d^3p_j}{(2\pi)^3 2E_j} (1 \pm f_j)\right) \\\nonumber
     &&\qquad\qquad\qquad\times (2\pi)^4\delta^4(p_1 + p_2 - p_3 - p_4)C_A \braket{|\mathcal{M}|^2},
\end{eqnarray}
where $f_i$ is the phase-space distribution function of the respective particle, $C_A$ a combinatorial factor which is 1/2 for identical initial state particles and 1 otherwise. Finally, $\braket{|\mathcal{M}^2|}$ is the squared amplitude summed over the initial and final spin states. Note that, as we assume that the initial population of the dark sector particles is negligible, dark Primakov processes such as $\phi + f\rightarrow \gamma' + f$,  or $\gamma' \rightarrow \gamma\phi$ decays, do not enter in Eqs.~\eqref{boltzeq1} and \eqref{boltzeq2}\footnote{We have checked that processes such as $\gamma^*\rightarrow \gamma'\phi$ are subleading in the parameter space we are interested in.}. However, after the FI process, we consider the $\gamma'$ decay in section~\ref{sec:cosmo}.

In the following, we describe the freeze-in analysis in two regimes:
\begin{enumerate}
    \item Strong freeze-in~\cite{Cosme:2023xpa}: $\TRH$ is smaller than the dark state masses. This requires relatively strong couplings to overcome the Boltzmann suppression.
    \item Standard high reheating temperature freeze-in: $\TRH$ is (well) above the masses of the two dark sector particles and SM masses.
\end{enumerate}

\subsection{Strong freeze-in}\label{strong_fi}
In this regime, we have $\TRH$ below one or both of the two dark state masses. The overall effect of this is that a Boltzmann suppression arises in the collision terms relevant for the calculation of the abundances of both dark sector particles, implying a drastic change in the viable parameter space to obtain the correct relic abundance~\cite{Cosme:2023xpa}.

To understand the essential behavior, let us consider the situation where the dark state masses are well below the EW scale, such that the leading FI processes involve only SM fermions and photons (in the numerical calculation using \micro\ the contributions from the other SM bosons are also taken into account). Using this simplification, Eqs.~\eqref{boltzeq1} and \eqref{boltzeq2} can be recast by using the yields $Y = n/s$, with $s = \frac{2\pi^2}{45}g_{*s}T^3$ the entropy density, resulting in 
\begin{eqnarray}\label{yield_strong_a}
     \frac{dY_{\phi}}{dT}  &=& -\frac{1}{\overline{H} T s}\left(\sum_f C_{f\bar{f}\rightarrow \phi\gamma'} + C_{\gamma\gamma\rightarrow \phi\phi} \right),\\
       \label{yield_strong_b}\frac{dY_{\gamma'}}{dT}  &=& -\frac{1}{\overline{H} T s}\left(\sum_f C_{f\bar{f}\rightarrow \phi\gamma'} + C_{\gamma\gamma\rightarrow \gamma'\gamma'}\right),
\end{eqnarray}
where $H/\overline{H} \equiv 1 + \frac{T}{3}\frac{d\log g_{*s}}{dT}$. The total yield produced between $\TRH$ and an arbitrary low temperature $T$ can be obtained by integrating Eqs.~\eqref{yield_strong_a} and~\eqref{yield_strong_b} over this temperature interval
\begin{eqnarray}\label{yield_strong_1}
     Y_{\phi(\gamma')}^\infty(T)  = Y_{\phi(\gamma')}(\TRH) - \int_{\TRH}^{T}dT' \frac{1}{\overline{H} T' s} C(T')\end{eqnarray}
 where $C(T') = \sum_f C_{f\bar{f}\rightarrow \phi\gamma'} + C_{\gamma\gamma\rightarrow \phi\phi(\gamma'\gamma')}$. As $T$ is much smaller than $m_{\gamma'}$ or $(m_\phi, m_{\gamma'})$, only the thermal tail of the distribution of initial states has the necessary energy to produce the dark states. In this way, as $\sqrt{s} \gg T$, the Maxwell-Boltzmann (MB) distribution becomes a good approximation to obtain the average in each collision term. For instance, for the process $f(p_1) + \bar{f}(p_2)\rightarrow \phi (k_1) + \gamma'(k_2)$, we have that the collision term becomes \cite{Gondolo:1990dk}
\begin{eqnarray}\label{collision_exact}
    C_{f\bar{f}\rightarrow \phi\gamma'}(T') &= & n_{f,e}^2 \braket{\sigma v}_{f\bar{f}\rightarrow \phi\gamma'} \\ \label{collision_exact2}
    &=&\int_{(m_\phi + m_{\gamma'})^2}^\infty d^3p_1d^3p_2 (\sigma v) e^{-E_1/T'}e^{-E_2/T'}\\ &=& \label{ys}\frac{2\pi^2T'}{(2\pi)^6}\int_{(m_\phi + m_{\gamma'})^2}^\infty ds \sigma \sqrt{s}(s - 4m_f^2)K_1\left(\frac{\sqrt{s}}{T'}\right) \,.
\end{eqnarray}
From the integrand of Eq.~\eqref{ys}, the exponential suppression is explicit if we recognize $T' = \TRH \ll \sqrt{s}$, since $K_1(\sqrt{s}/T') \propto e^{-\sqrt{s}/T'}$. An equivalent result is obtained for $C_{\gamma\gamma\rightarrow \phi\gamma'}$.

Let us now follow an even simpler argument, which allows us to obtain analytical expressions for the collision terms. Let us take the case $\TRH < (m_\phi, m_{\gamma'})$, such that the dark states behave as non-relativistic particles. We use detailed balance (see also \cite{Bringmann:2021sth}) in order to calculate the collision terms appearing in Eqs.~\eqref{yield_strong_1} and \eqref{yield_strong_2}. We find
\begin{eqnarray}\label{db1}
   C_{f\bar{f}\rightarrow \phi\gamma'} =  \braket{\sigma v}_{f\bar{f}\rightarrow \phi\gamma'} n_{f,e}^2 = \braket{\sigma v}_{\phi\gamma' \rightarrow f\bar{f}} n_{\phi,e}n_{\gamma',e} \,.
\end{eqnarray}
The average annihilation cross section on the RHS of \eqref{db1} is obtained by expanding $\sigma v$ in powers of velocities, retaining the leading term. We have used \calchep \cite{Belyaev:2012qa}, and the corresponding expressions for each final SM fermion final state are given in App.~\ref{app_strongfi1}. As we are in the non-relativistic limit, the equilibrium number densities in the RHS of \eqref{db1} are $n_{\phi(\gamma'),e} \approx g_{\phi(\gamma')}\left(\frac{m_{\phi(\gamma')} T}{2\pi}\right)^{3/2}e^{-m_{\phi(\gamma')}/T}$.  
On the other hand, and equivalently to the previous case, for the photon-photon annihilation into dark states, we have that 
\begin{eqnarray}\label{db2}
C_{\gamma\gamma\rightarrow\phi\phi(\gamma'\gamma')} = \braket{\sigma v}_{\gamma\gamma\rightarrow\phi\phi(\gamma'\gamma')} n_{\gamma,e}^2 = \braket{\sigma v}_{\phi\phi(\gamma'\gamma')  \rightarrow\gamma\gamma}n_{\phi(\gamma'),e}^2 \,.
\end{eqnarray}
The average annihilation cross sections times velocity on the RHS of \eqref{db2} are obtained in the same way as for $\braket{\sigma v}_{\phi\gamma' \rightarrow f\bar{f}}$, but this time the expressions contain the $s$-wave term, 
\begin{eqnarray}\label{sigv_1100}
    \braket{\sigma v}_{\phi\phi\rightarrow \gamma\gamma} &\approx& \frac{2\Gg^4 m_{\phi}^6}{\pi (m_\phi^2 + m_{\gamma'}^2)^2} + \mathcal{O}(v^2) \\\braket{\sigma v}_{\gamma'\gamma'\rightarrow \gamma\gamma} &\approx& \frac{\Gg^4 m_{\gamma'}^6}{3\pi (m_\phi^2 + m_{\gamma'}^2)^2} + \mathcal{O}(v^2)\,.
\end{eqnarray}
In this way, each collision term entering in Eq.~\eqref{yield_strong_1} is written algebraically, and we only need to integrate over the temperature to obtain the corresponding yields. 

Neglecting the yields at $\TRH$, and after integrating out the collision terms in Eq.~\eqref{yield_strong_1}, the resulting individual yield can be written as
\begin{eqnarray}\label{yield_strong_2}
     Y_{\phi}  &=& Y_{\gamma\gamma\rightarrow\phi\phi} + Y_{f\bar{f}\rightarrow\phi\gamma'} \\
     Y_{\gamma'}  &=& Y_{\gamma\gamma\rightarrow\gamma'\gamma'} + Y_{f\bar{f}\rightarrow\phi\gamma'}
     \label{eq:gamma_yield}
     \end{eqnarray}
where 
\begin{eqnarray}
\label{analytic_1}Y_{\gamma\gamma\rightarrow\phi\phi} &=& \frac{135\sqrt{5} g_\phi^2 \Gg^4 m_\phi^7 M_P}{16 \pi^7 g_{*s}\sqrt{2g_*}(m_\phi^2 + m_{\gamma'}^2)^2} \frac{(2m_\phi + \TRH)}{\TRH}e^{-{2m_\phi}/{\TRH}} \\ \label{analytic_2}
Y_{\gamma\gamma\rightarrow\gamma'\gamma'} &=& \frac{45\sqrt{5} g_{\gamma'}^2 \Gg^4 m_{\gamma'}^7 M_P}{32 \pi^7 g_{*s}\sqrt{2g_*}(m_\phi^2 + m_{\gamma'}^2)^2} \frac{(2m_{\gamma'} + \TRH)}{\TRH}e^{-{2m_{\gamma'}}/{\TRH}}  \\ \label{analytic_3}
Y_{f\bar{f}\rightarrow\phi\gamma'} &=& \frac{45\sqrt{90} g_\phi g_{\gamma'} M_P}{16 \pi^6 g_{*s}\sqrt{g_*}} \frac{(m_\phi m_{\gamma'})^{3/2}}{m_\phi + m_{\gamma'}} \braket{\sigma v}_{eff} e^{-(m_\phi + m_{\gamma'})/T_{RH}} 
\end{eqnarray}
with 
\begin{eqnarray}
    \braket{\sigma v}_{eff} = \frac{1}{T_{RH}}\sum_{f} \braket{\sigma v}_{\phi\gamma'\rightarrow
     f\bar{f}}(T_{RH})\,.
\end{eqnarray}
All terms in the last expression are given explicitly in App.~\ref{app_strongfi1}, and the sum runs only over those fermions which fulfill the condition $m_\phi + m_{\gamma'} > 2m_f$.

\begin{figure}[t!]
    \centering
\includegraphics[scale=0.43]{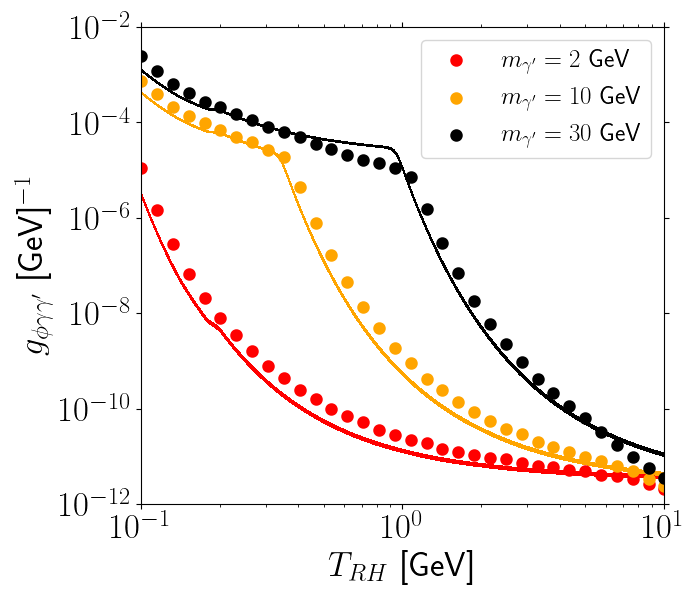}
\includegraphics[scale=0.6]{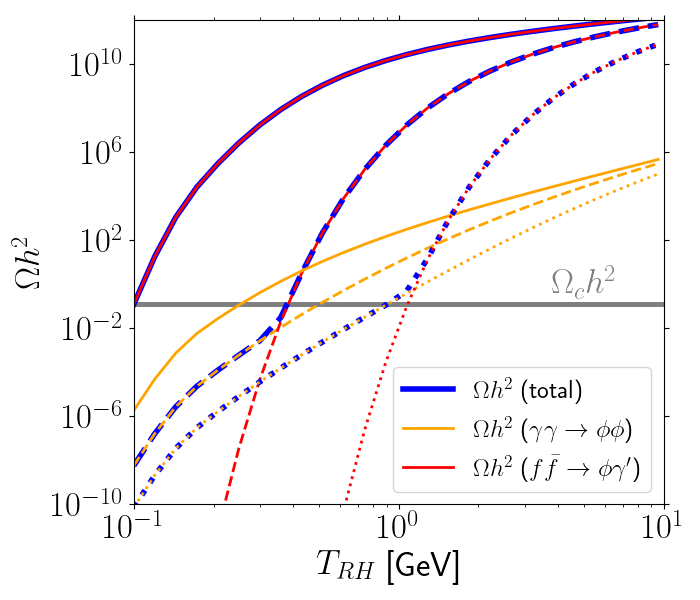} \caption{(left) Comparison between \micro\ (dots) and the analytic relic abundance in the strong FI (solid lines), considering $m_\phi = 1$ GeV, with the all the points and curves fulfilling the correct relic abundance. (right) Relic abundance in the strong FI regime. In each case, we show the total relic abundance for $m_\phi = 1$ GeV, $m_{\gamma'} = 2$ GeV (solid blue), 10 GeV (dashed blue), 30 GeV (dotted blue), for $\Gg = 10^{-5}$ GeV$^{-1}$. The corresponding set of orange and red curves correspond to the relic abundance contribution from $\gamma\gamma$ and $f\bar{f}$ annihilation processes, respectively. All the results were obtained with \micro\ .}
    \label{fig:relic_lowTR}
\end{figure}

Plugging the set of yields \eqref{analytic_1}, \eqref{analytic_2} and \eqref{analytic_3} into the formula for the parameter density \eqref{par_den}, we can compare our analytic results with the numerical results obtained with \micro. In Fig.~\ref{fig:relic_lowTR}~(left) we show the contours in the parameter space which fulfill the correct relic abundance, setting $m_\phi = 1$ GeV and varying $m_{\gamma'}$. We observe a strong increase of $\Gg$ as $\TRH$ decreases as a result of the Boltzmann suppression in collision terms, with a stronger effect as $m_{\gamma'}$ increases. In particular, the kink in each curve is due to the change of the leading FI process: In each curve, from left to the kink, $\gamma\gamma\to \phi\phi$ is the leading FI production channel, while from the kink to the right, $f\bar{f}\rightarrow\phi\gamma'$ is the leading process. The small differences between the analytic and numerical methods are due to the fact that in the former, we truncate the average annihilation cross-section expansion in the leading partial wave, whereas the numerical method integrates over the whole MB momentum distribution of the initial particles. 

In Fig.~\ref{fig:relic_lowTR}~(right), we show the contribution of the different scattering channels  to the FI, where we use the same masses as in the plot on the left. Here,  it is easier to see that there is a low threshold temperature that separates the leading channel that contributes to the relic abundance of the DM candidate, showing a clear change in the behavior of the relic abundance as a function of $T_{RH}$. Notice that, it can also occur that the exponential suppression in total relic abundance occurs for $T_{RH} > m_\phi$, as shown by the blue dotted curve. In this case, it is necessary that $T_{RH} < m_{\gamma'}$. 

 \subsubsection*{Thermalization in the strong freeze-in regime}
\begin{figure}[t!]
    \centering
\includegraphics[scale=0.35]{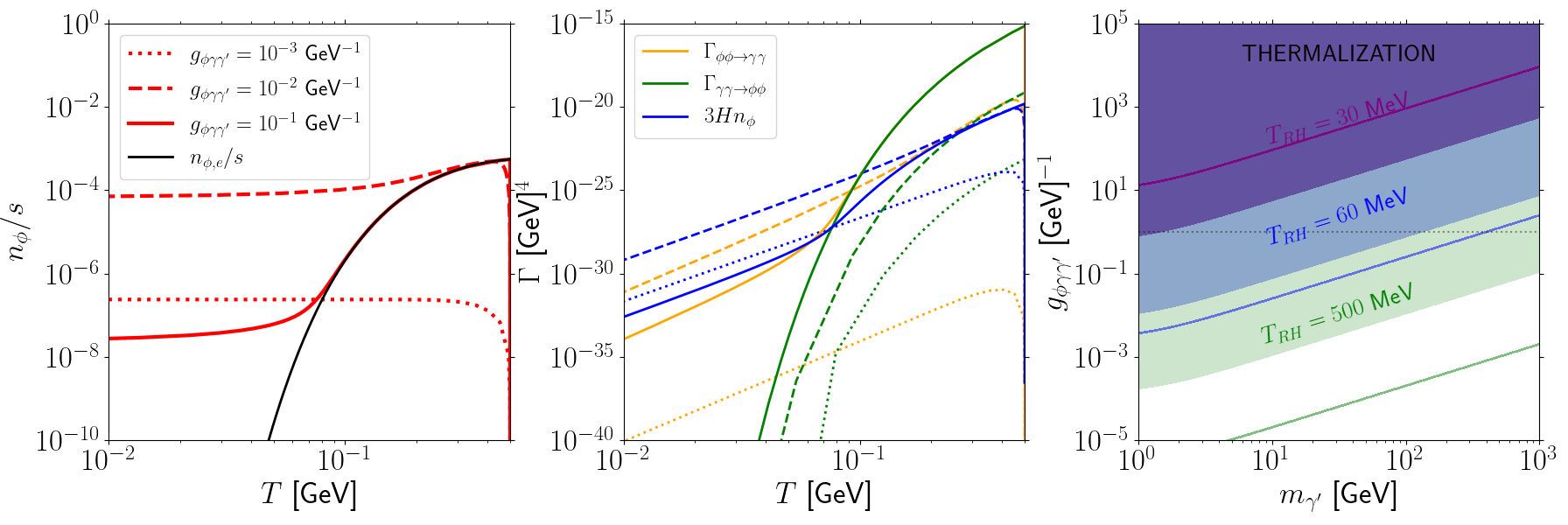}
\includegraphics[scale=0.34]{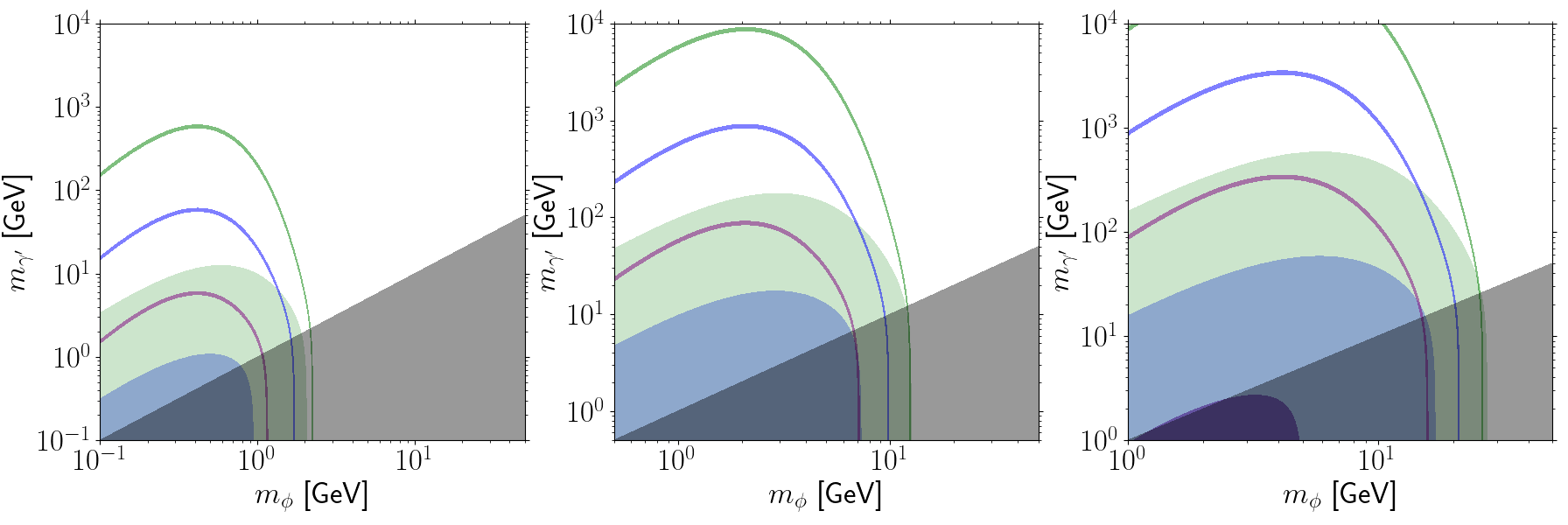}
    \caption{(above) In the plot in the left we show the yield evolution resulting from the numerical solution of Eq.~\eqref{simple_beq}, considering $m_\phi = 1$ GeV, $m_{\gamma'} = 30$ GeV and $\TRH = 0.5$ GeV. The plot in the middle shows the density rates in the strong FI regime versus the Hubble term in the Boltzmann equation, with the solid, dashed and dotted lines following the same values shown in the plot in the left. In the plot in the right we show the thermalization condition $Y_{\gamma\gamma} > Y_{\phi,e} (\TRH)$ with a specific color region depicting a particular reheating temperature. The purple, blue and green lines give the correct relic abundance assuming the corresponding value of $\TRH$ shown in each region. The dotted horizontal line indicates $m_\phi = \Gg^{-1}$, where the EFT treatment becomes questionable. (below) The thermalization condition is marked by the color regions. From left to the right we set $T_{RH} = 0.1, 0.5$ and 1 GeV. The corresponding purple, blue and green lines are the contours of correct relic abundance for $\Gg = 10^{-4}, 10^{-3}$ and $10^{-2}$ GeV$^{-1}$, respectively.}
    \label{fig:thermali}
\end{figure}

Due to the exponentially increasing behavior of $\Gg$ in strong FI scenarios, one may wonder about the possibility of thermalization between the visible and dark sectors, i.e. forward and backward particle reactions involving dark states equilibrate. To estimate whether this happens, we consider the extreme case in the strong FI regime where $\gamma\gamma\rightarrow\phi\phi$ dominates the production of relic abundance and the population of DP is negligible as a result of Boltzmann suppression. In this way, following a similar treatment as \cite{Cosme:2023xpa}, we can simply describe the abundance of ALPs via
\begin{eqnarray}\label{simple_beq}
    \dot{n}_\phi + 3Hn_\phi = \Gamma_{\gamma\gamma \rightarrow \phi\phi} - \Gamma_{\phi\phi\rightarrow\gamma\gamma }
\end{eqnarray}
 where
 \begin{eqnarray} \label{forward}
    \Gamma_{\gamma\gamma \rightarrow \phi\phi} &=& n_{\gamma,e}^2 \braket{\sigma v}_{\gamma\gamma \rightarrow \phi\phi} = n_{\phi, e}^2 \braket{\sigma v}_{\phi\phi\rightarrow\gamma\gamma  } \\
    \label{backward}\Gamma_{\phi\phi\rightarrow\gamma\gamma } &=& n_{\phi}^2 \braket{\sigma v}_{\phi\phi\rightarrow\gamma\gamma} 
 \end{eqnarray}
and $\braket{\sigma v}_{\phi\phi\rightarrow\gamma\gamma} $ is given in \eqref{sigv_1100}. Typically, in FI scenarios, we have $n_\phi \ll n_{\phi,e}$ at $T=\TRH$. However, if $\Gg$ takes too large values, it can happen that the inverse reaction \eqref{backward} can be equilibrated with the forward reaction \eqref{forward}. We show the numerical result of solving Eq.~\eqref{simple_beq} for $n_\phi(\TRH) \approx 0$ in Fig.~\ref{fig:thermali}~(upper left) for different values of $\Gg$ and fixed masses of the dark states. Once $\Gg$ takes too high values (solid and dashed red), the two rates equilibrate for a period and then $\phi$ freezes out. Equivalently, in Fig.~\ref{fig:thermali} (upper middle) we show the evolution of each reaction density compared to the term of the Boltzmann equation $3Hn_\phi$, where we observe that $\Gamma_{\phi\phi\rightarrow\gamma\gamma}$ becomes equilibrated with its inverse reaction $\Gamma_{\gamma\gamma\rightarrow \phi\phi}$ for very high $\Gg$. As the value decreases $\Gg$, the former reaction cannot overcome the Hubble term or be equilibrated with its reverse reaction. 

In this strong FI regime, a simple thermalization criterion is $Y_{\gamma\gamma} > Y_{\phi,e}(\TRH)$. It turns out that none of the parameter spaces fulfills this condition, since this is achieved only for $\Gg > m_\phi^{-1}$, which is outside the validity of our EFT description. We exemplify this fact with a few benchmark points in Fig.~\ref{fig:thermali}.

\subsection{Standard Freeze-In}
We now take $\TRH$ to be very high such that we can neglect all dark sector and SM masses. In this case, the channel $f\bar{f}\rightarrow \phi \gamma'$ becomes the leading channel for FI production. Let us obtain the yield of the corresponding process. 

The Boltzmann equation for the yield of the ALP and an identical for the DP can be written as (see Eq.~\eqref{yield_strong_1}), 
\begin{eqnarray}\label{eq3}
 \frac{dY_\phi}{dT} &=& - \frac{45}{2\pi^3 g_{*s}}\sqrt{\frac{90}{g_{*s}}}\frac{M_P}{T^6}\sum_f C(f\bar{f}\rightarrow \phi\gamma'),
\end{eqnarray}
where $M_P$ is the Planck mass, and where we have considered $g_{*s}$ constant. The collision term is given by $C(f\bar{f}\rightarrow \phi {\gamma'})  = \braket{\sigma_ {f\bar{f} \rightarrow\phi\gamma'}v}n_{f,e}^2$, and considering that the number density of relativistic fermions is given by $n_{f,e} = g_f\frac{\zeta(3)}{\pi^2}T^3$, with $g_f$ the internal degrees of freedom of the fermion. Eq.~\eqref{eq3} can be written as  
\begin{eqnarray}
    \frac{dY_\phi}{dT} &=& -\frac{45\zeta(3)^2 g_{f}^2 M_P}{2\pi^7 g_{*s}}\sqrt{\frac{90}{g_{*s}}} \sum_f\braket{\sigma_{f\bar{f} \rightarrow \phi\gamma'}v}. 
\end{eqnarray}
Assuming that the effective degrees of freedom in energy and entropy remain constant, the yield can then be integrated to get \cite{Bernal:2019mhf}
\begin{eqnarray}\label{fi_formula}
    Y_\phi(T) = Y_\phi(\TRH) - \frac{45\zeta(3)^2 g_f^2 M_P}{2\pi^7 g_{*s}}\sqrt{\frac{90}{g_{*s}}}\sum_f\int_{\TRH}^T dT'\braket{\sigma_{f\bar{f} \rightarrow \phi\gamma'}v}(T') \,.
\end{eqnarray}
At high energies, and similarly to the strong FI regime, we have different contributions to the average annihilation cross-section times velocity, depending on the flavor of the fermion (for details see App~\ref{app_A1}). Summing all the contributions we have,  
\begin{eqnarray}
    \sum_f\braket{\sigma_{f\bar{f} \rightarrow \phi\gamma'} v} &=& \sum_{\text{SM neutrinos}} \frac{e^2\Gg^2}{24\pi c_W^2} + \sum_{\text{charged lep.}} \frac{5 e^2\Gg^2}{96\pi c_W^2} \\ 
    &&\qquad\qquad+ \sum_{\text{up-type quarks}} \frac{17 e^2\Gg^2}{2592\pi c_W^2} + \sum_{\text{down-type quarks}} \frac{5 e^2\Gg^2}{2592\pi c_W^2}  \\
    &=& \frac{265 e^2\Gg^2}{864\pi c_W^2},
\end{eqnarray}
where $c_W$ is the cosine of the Weinberg angle. Assuming $Y_\phi(\TRH) \approx 0$, the final yield is then given by
\begin{eqnarray}\label{fi_formula_b}
    Y_{\phi}^\infty = \frac{1325\zeta(3)^2 g_f^2 e^2 \Gg^2 M_P}{192\pi^8 c_W^2 \gss}\sqrt{\frac{90}{g_{\star}}}\TRH\, .
    \label{eq:initial_yield}
\end{eqnarray}
Again, recall that $Y_{\gamma'}^\infty = Y_\phi^\infty$. 

\begin{figure}
	\centering 
\includegraphics[scale=0.5]{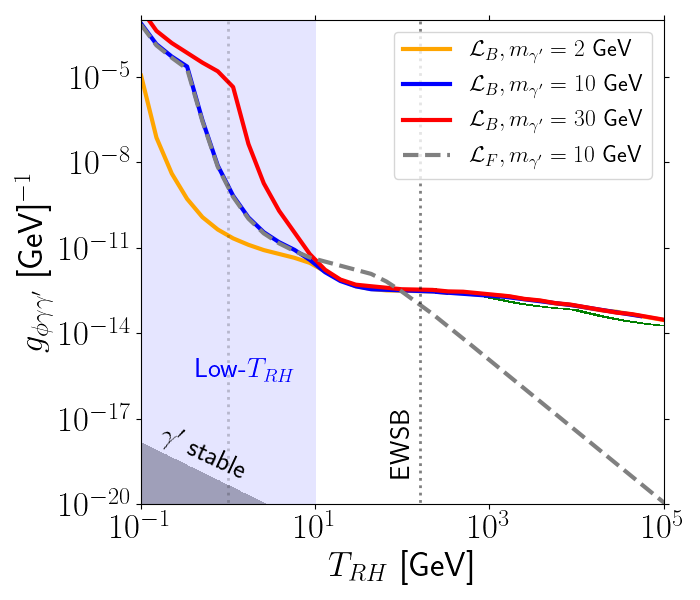}
\caption{Parameter space fulfilling the correct relic abundance via freeze-in for $m_\phi = 1$ GeV. The solid colored lines are obtained by making use of $\mathcal{L}_B$ EFT, whereas the dashed gray with $\mathcal{L}_F$, all of them obtained with \micro . The solid green line is the result based on the analytic estimation given in Eq.~\eqref{fi_formula_b} which is valid for high temperatures and independent of the DP mass.  }
\label{fig:fi_final}
 \end{figure}

In Fig.~\ref{fig:fi_final}, we show the corresponding values in parameter space that satisfy the correct relic abundance (colored lines). For the moment, we have ignored the contribution from the DP decay to the ALP abundance. This will be covered in the next section. In particular, at large $\TRH$, the three solid lines converge independently on $m_{\gamma'}$, with a slow change in $\Gg$ as $\TRH$ increases, as suggested by Eq.~\eqref{fi_formula_b}. Furthermore, the blue region shows the strong freeze-in regime, with a strong dependence on $m_\gamma'$.
 
Let us briefly discuss some noteworthy features visible in the $(m_\phi, m_{\gamma'}) < \TRH \lesssim v_{EW}$ region of Fig.~\ref{fig:fi_final} (with $v_{EW} = 256$ GeV the electroweak scale). See also Appendix~\ref{app:unitarity} for additional discussion. First, for $\TRH \sim m_W$, the annihilation channel $W^+W^-\rightarrow \phi\gamma'$ contributes in a subleading way to the FI of the dark states. However, the $Z$ boson in the $s$-channel becomes relevant as an intermediate state because it recovers $SU(2)_L\times U(1)_Y$ gauge symmetry, and ensures that the amplitudes do not grow too fast at high energies.  That is, in scenario $\mathcal L_F$, the process $W^+W^-\rightarrow\phi\gamma'$ mediated only by the SM photon grows indefinitely with the energy or $\TRH$, so that $\Gg$ must decrease to compensate for this growth amplitude (dashed gray line in Fig.~\ref{fig:fi_final}), while in EFT $\mathcal L_B$, this process is energetically under control since the gauge symmetry is recovered. In this way, in the $\mathcal L_B$ EFT, fermion-antifermion annihilation remains the leading process of production for temperatures below the EWSB scale, although $W^+W^-$ makes a small contribution to the relic abundance with less than 1\% to the relic abundance, and $Zh$ annihilation to a lesser extent. Notice that for $\TRH$ below the EWSB scale, there are less than $\mathcal{O}(10)$ differences for $\Gg$ for a fixed $\TRH$ between $\mathcal{L}_F$ and $\mathcal{L}_B$, due to the presence of the $Z$ boson in the latter case.

\subsubsection*{Thermalization in the standard freeze-in regime}
As in the strong freeze-in scenario, high couplings could lead to thermalization of the dark particles with the thermal bath and undergo a freeze-out process, instead of freeze-in. Let us quickly estimate when this would be the case. For simplicity, we consider only the processes with charged fermions. We define the interaction rate with charged fermions as
\begin{align}
    \Gamma_q= n_q \langle\sigma_{q\bar q\rightarrow \phi\gamma'}v\rangle.
\end{align}
{Here,} $n_q=\sum_i Q^2_i n_i=(\zeta(3)/\pi^2)g_q(T) T^3$ and $Q_i$ is the charge of the i-th particle and $g_q$ the effective number of relativistic degrees of freedom of charge particles.  We now compare the rate to the Hubble parameter, finding that the condition for thermalization is
\begin{align}
    \Gg\gtrsim 8\times 10^{-16}\, \mbox{GeV}^{-1}\, \sqrt{\frac{10^{14}\,\mbox{GeV}}{g_q T_{RH}}}.
    \label{eq:thermalization_stdFI}
\end{align}
As we will see in the next section the standard freeze-in mechanism is safe from thermalization in all the relevant parameter space.

In summary, the FI production of both dark states is viable, for high and low $\TRH$. These two cases differ strongly in the required values of $\Gg$ to fulfill the correct relic abundance. This fact has profound consequences not only for the relevant constraints in each case, but also for the possible way to test each scenario. In the following section we explore these features in more detail.

\section{Dark matter evolution in the early universe}\label{sec:cosmo}
The cosmological evolution of our model depends strongly on the lifetime of the dark photon and the maximum temperature in the universe which, for simplicity, we assume is the reheating temperature, $\TRH$\footnote{We work in the instantaneous reheating approximation, but due to the most-likely intricate nature of the inflationary process, the maximum temperature in the universe could be different than $\TRH$, see for instance \cite{Giudice:2000ex,PhysRevD.101.123507}.}.  The decay rate (in vacuum) of the process $\gamma'\rightarrow \gamma+\phi$ is given by \cite{Arias:2020tzl}
\begin{eqnarray}
\Gamma_{\gamma'\rightarrow \gamma + \phi} = \frac{\Gg^2m_{\gamma'}^3}{96\pi}\left(1 - \frac{m_\phi^2}{m_{\gamma'}^2}\right)^3.
\label{eq:decay_rate}
\end{eqnarray}
{Notably, the decay typically happens (see below) when the DP is already non-relativistic, $T_{\rm decay}\lesssim m_{\gamma'}$ and therefore the lifetime does not feature an additional $\gamma$-factor, $\tau=(\Gamma_{\gamma'\rightarrow \gamma + \phi})^{-1}$.}

A stable dark photon is obtained if the lifetime is larger than the age of the universe. For DP heavier than an MeV this happens only for fairly small couplings (leaving aside the fine-tuned case of nearly degenerate masses). 

In this work we are interested in the regime where the DP is unstable, leaving the ALP as the only DM candidate. The decay condition is given by,
\begin{align}
    H(T_{\rm decay})=\Gamma_{\gamma'\rightarrow \gamma\phi}.
\end{align}
Assuming a radiation-dominated universe, the decay temperature is given by
\begin{eqnarray}
    T_{\rm decay}&\approx& \frac{1}{10}\Gg \sqrt{\frac{M_P \left(\mgamma^2-m_\phi^2\right)^3}{(\gs(T_{\rm decay}))^{1/2} \mgamma^3}}
    \\\nonumber
    &\approx& 3.5\, \mbox{MeV} \left(\frac{\Gg}{10^{-15}\,\mbox{GeV}^{-1}}\right)\left(\frac{10}{g_{*}(T_{\rm decay})}\right)^{1/4}\left(\frac{m_{\gamma'}}{1000\,\mbox{GeV}}\right)^{3/2},
\label{eq:decay_temp}
\end{eqnarray}
where, in the {second line}, we have taken  $m_{\phi}\leq 0.1\, m_{\gamma'}$ and $M_P=2.4\times 10^{18}$~GeV is the reduced Planck mass. 
An early decay of the DP requires high couplings and/or high masses, which could conflict with the freeze-in requirement on the coupling. Also, large DP masses lead to high initial momentum for the ALP population from the decay, potentially leading to dangerous interference with structure formation. We will come back to this point later.   
A mass $\mgamma \sim$10~MeV together with $\Gg \approx 10^{-15}$~GeV$^{-1}$, leads to {a} decay after matter-radiation equality at $T_{\rm decay} \approx 0.1$~eV, also ruling out the option of the ALP to be a viable DM candidate.

The freeze-in production mechanism is sensitive to the hierarchy between dark particle masses and the reheating temperature, as discussed in the previous section. In the following, we will explore the cosmological implications of our model and highlight the strongest constraints from the successful evolution of the early universe. We will then delineate the viable parameter space for the model under both the standard and strong freeze-in mechanisms. For the latter, we will address the subtleties of the cosmological evolution specific to that scenario.

\subsection{Including the decay contribution}

The dynamics of the decay are governed by the following Boltzmann equations
\begin{subequations}
\begin{eqnarray}
    \frac{d N_{\gamma'}}{da}&=&-\Gamma_{\gamma'\rightarrow\gamma\phi}\frac{N_{\gamma'}}{Ha
},\\
\frac{d N_{\phi}}{da}&=&\Gamma_{\gamma'\rightarrow\gamma\phi}\frac{N_{\gamma'}}{Ha
},\\
\frac{dT}{da}&=&-\frac{T}a+\frac{15}{4\pi^2}\frac{\Gamma_{\gamma'\rightarrow\gamma\phi}}{H a^4}\frac{N_{\gamma'}\mgamma}{T^3},
\end{eqnarray}
\label{eq:BE_decay}
\end{subequations}
with $a$ the scale factor and $T$ the photon {temperature. Moreover,} we have assumed a non-relativistic dark photon. We solve the set of equations using the Hubble parameter
\begin{align}
H=\sqrt{\frac{\rho_r+\rho_{\gamma'}}{3 M_P^2}}.
\end{align}
Here, the total radiation density is given by $\rho_r=\frac{\pi^2}{30}\gs(T)T^4$ and we have neglected the contribution from the ALP because we will require it to satisfy the observed abundance today, making it subdominant at early times. For the DP energy density, we use $\rho_{\gamma'}=\frac{N_\gamma}{a^3}\mgamma$. The initial number of DPs and ALPs is obtained from the yield of Eq.~\eqref{eq:initial_yield}, that is, $N^i_{\gamma',\phi}= Y_{\phi}^\infty S_i$, with $S_i=S(T_i)$ the entropy evaluated at some initial temperature $T_i$ where we start our integration. As a general case, we will consider the decay to occur while the DP is non-relativistic. This is certainly a good approach for most of the parameter space in which we are interested. Thus, the initial integration temperature is chosen to satisfy $T_i\lesssim \mgamma$. 

It is possible to find analytical solutions to the above equations assuming that DP never dominates the energy density (which is an excellent approximation in the allowed parameter space). They are
\begin{eqnarray}
    N_{\gamma'}&=&N_{\gamma' }^{i}\, e^{-a^2/2 a_{\rm decay}^2} ,\label{eq:N_DP}\\
    N_\phi&=&N_\phi^i+{N_\gamma^i}(1-e^{-a^2/2a_{\rm decay}^2}),
    \label{eq:N_alp}
\end{eqnarray}
where $a_{\rm decay}$ is the scale factor at the decay. Note that, as expected, the final number of ALPs after decay is $N_\phi^f=N_\phi^i+N_{\gamma'}^i=2N_\phi^i$, because the initial yield for DPs and ALPs is the same, see Eq.~\eqref{eq:initial_yield}. On the other hand, the expression for the temperature evolution is more involved, but it can be greatly simplified if computed by parts and at first order in the parameter $\frac{\Gamma_{\gamma'\rightarrow \gamma+\phi}}{2H(a_i)}$. Before the decay, it is given by
\begin{align}
    \frac{T(a)}{T_i}= \,\frac{a_i}{a}\left(1+\frac{\rho_{\gamma',i}}{\rho_{\gamma,i}}\left(\frac{a}{a_i}\right)^3\frac{\Gamma}{6H(a_i)}\right)^{1/4}.
\end{align}
{Here,} $\rho_{\gamma,i},\rho_{\gamma',i}$ are the energy densities of the photon and DP at $T=T_i$, respectively. After the `reheating' of the DP into photons and ALPs, the temperature recovers the usual behavior
\begin{align}
    \frac{T(a)}{ T_{\rm end}}=\frac{a_{\rm end}}{a},
\end{align}
with $T_{\rm end}$ and $a_{\rm end}$ the temperature and scale factor once the decay has completely ended\footnote{In the sudden decay approximation this temperature coincides with the decay temperature.}. 
It is possible to quantify the entropy injection of the DP into photons by imposing energy conservation just before and just after the decay (ignoring the contribution of the ALP, as it should be subdominant),
\begin{align}
    \rho_{\gamma}(T_i) \left(\frac{T_{\rm end}}{T_i}\right)^4=\rho_{\gamma}(T_i) \left(\frac{\Tdec}{T_i}\right)^4+\rho_{\gamma'}(T_i) \frac{\gs(\Tdec)}{\gs(T_i)} \left(\frac{\Tdec}{T_i}\right)^3,
\end{align}
leading to
\begin{align}
   \left( \frac{T_{\rm end}}{\Tdec}\right)^3= \left[1+\frac{\rho_{\gamma'}(T_i)}{\rho_{\gamma}(T_i)}\frac{\gs(\Tdec)}{\gs(T_i)} \frac{T_i}{\Tdec}\right]^{3/4}.
\end{align}
To conclude this part, we derive the corrected expression for the relic abundance of the ALP. In effect, the decay transfers part of the energy of the dark sector into the visible one, therefore, even though the ALP number is doubled with respect to the initial one, it is not necessarily what happens between the initial and final yield. The abundance today is given by
\begin{align}
\Omega_{\phi,0}=Y_\phi^f s(T_0)\frac{m_\phi}{\rho_{c,0}}.
\end{align}
$Y_\phi^f$ is the final (after the decay) ALP yield. Using $Y=N/S$ we find
\begin{align}
Y_\phi^f=\frac{N_\phi^f}{S_{\rm end}}=2{Y_{\phi}^\infty} \frac{S_{\rm decay}}{S_{\rm end}},
\end{align}
where we have used that, initially, the ALP and DP total number are the same and that entropy is conserved just before the decay, that is ${Y_{\phi}^\infty} S_{RH}={Y_{\phi}^\infty} S_{\rm decay}$. {Inserting this into the expression for the abundance}, 
\begin{align}
    \Omega_{\phi,0}=2{Y_{\phi}^\infty}\, s(T_0) \,\frac{m_\phi}{\rho_{c,0}} \frac{S_{\rm decay}}{S_{\rm end}}.
    \label{eq:DM_relic}
\end{align}

\begin{figure}[t]
    \centering
    \includegraphics[width=1\columnwidth]{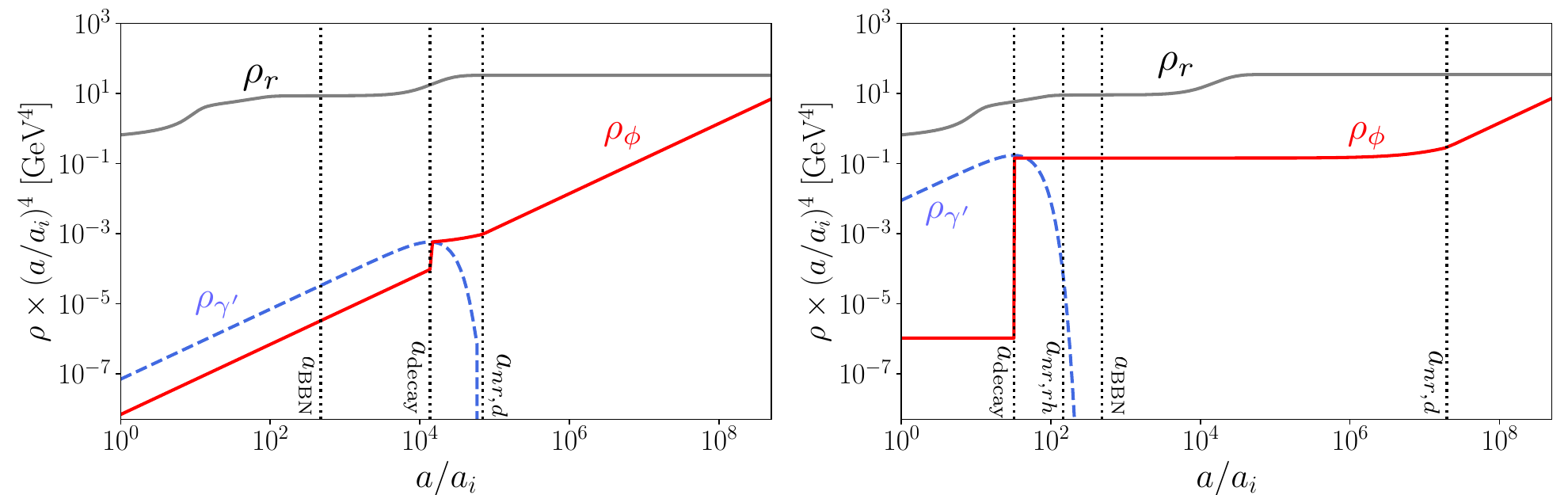}
    \caption{Evolution of the energy density for the SM radiation (gray), dark photon (dashed blue) and the axion-like particle (solid red) as a function of the scale factor. Relevant moments of the evolution are highlighted in dotted vertical lines (see text for details). For the left plot, the ALP and the DP are non relativistic from the beginning of the integration at 1 GeV of the Boltzmann equations. The plot of the right starts integration with a non-relativistic dark photon and a relativistic ALP.  }
    \label{fig:energy_evolution}
\end{figure}

In Fig.~\ref{fig:energy_evolution} we present the evolution of the energy density for the DP (dashed blue), the ALP (solid red) and the radiation in the SM (solid gray) as a function of the scale factor (normalized to some initial arbitrary instant). We have obtained the evolution by numerically solving the Boltzmann equations \eqref{eq:BE_decay} at the initial temperature of 1 GeV. Both panels use $\TRH=10^{10}~$GeV and give the correct ALP abundance. For the left panel, the benchmark values are $\mgamma=7.5$~TeV, $m_\phi=740$~GeV and $\Gg=2.6\times 10^{-18}~$GeV$^{-1}$. For that panel, ALP and DP both start at the initial integration temperature as non-relativistic. At $a=\adec$ the DP decays, adding a relativistic population that has a higher energy than the existing non-relativistic one, therefore, we see it redshifts as radiation. At $a=a_{nr,d}$ the population from the decay becomes nonrelativistic, from which both populations redshift as matter. For the right panel, the benchmark values are $\mgamma=8.7$~TeV, $m_\phi=67$~MeV and $\Gg=8.7\times 10^{-16}~$GeV$^{-1}$. In this case, the ALP starts at the initial temperature as relativistic and the DP as non-relativistic. Before the DP decay, the reheating ALP population becomes non relativistic at $a=a_{nr,rh}$. Subsequently, the DP decays at $a=\adec$, corresponding to $\Tdec\sim 55$~MeV, introducing a corresponding relativistic ALP population. We also show the scale factor at $4$~MeV, which we label as $a_{\rm BBN}$, for reference. The ALP indeed contributes to the effective number of neutrinos at that epoch, but the energy density is too low to have an observable effect. Later, the ALP population from the decay becomes non-relativistic at $a_{ nr, d}$ from where the whole ALP population behaves as cold dark matter. 
%%%%

%%%%%
In the following, we will review cosmological constraints that shall be met in order not to interfere with observations.

\subsection{Warmness of the DM}

When the DP decays, the conservation of energy and momentum of the process fixes the magnitude of the momenta for the final states,
\bea
\mgamma &\approx &\sqrt{k_\phi^2+m_\phi^2}+k_\gamma, \,\,\,\,\,\,\,\,\,\mbox{and}\,\,\,\,\,\,\,
 k_\phi\sim k_\gamma=k_{\rm decay},\\
\Rightarrow k_{\rm decay}&=&\frac{\mgamma^2-m_\phi^2}{2\mgamma}.
\eea
Therefore, the ALPs from the decay are relativistic and their momentum redshifts according to $p_\phi=k_{\rm decay} \, (a_{\rm decay}/a)$. The time when they transition to being non-relativistic is determined by
\bea
p_{\phi,nr}\sim m_\phi \Rightarrow a_{nr}\approx \frac{\mgamma}{2m_\phi}a_{\rm decay}\,,
\eea
where we have taken $k_{\rm decay}\approx \mgamma/2$, {for} simplicity.
{After the decay-originated ALPs are non-relativistic their average velocity} is approximately given by
\be
\langle v_\phi\rangle\sim  (a_{nr}/a).
\ee
The ALP population shall be sufficiently cold before entering the matter dominated era and not interfere with structure formation, \cite{Viel:2005qj, Viel:2007mv,Baur:2015jsy,Chatrchyan:2020pzh}, satisfying \begin{align}
    \langle v\rangle \sim 10^{-3} c\,\,\,\,\, \mbox{ at}\,\, a=a_{\rm eq}. 
\end{align}
That is, the requirement translates into 
\be
\Tdec\gtrsim\,\frac{ \mgamma}{ m_\phi} \frac{T_{\rm eq}}{2\times10^{-3} }\approx 400 \, \mbox{eV} \left(\frac{ \mgamma}{ m_\phi} \right),
\label{eq:tdec_warm}
\ee
with $T_{\rm eq}\sim 0.8$ eV, the temperature at matter-radiation equality.
In terms of the coupling constant we find
\be
\Gg\gtrsim 2.6 \times 10^{-15}~\mbox{GeV}^{-1}\, \left(\frac{1\,\mbox{GeV}}{\mgamma}\right)^{3/2} ,
\ee
where we have taken $m_\phi=10^{-1}\,\mgamma$, and the constraint tightens for greater mass differences. 
Comparing the above expression with \eqref{eq:thermalization_stdFI}, we find that ensuring that the ALP population from decay remains sufficiently cold to avoid interfering with structure formation becomes challenging at high reheating temperatures and large DP masses. 

Let us remark that the first ALP population from freeze-in is produced with momentum $\sim \TRH$, therefore, it becomes nonrelativistic at temperatures of the order of the ALP mass.  

\subsection{Dark photon dominance and decay around BBN}
One potential scenario is that the DP itself dominates the energy density in the early universe, decaying before BBN \cite{Hannestad:2004px}. For a non-relativistic DP, {\it i.e.} $\mgamma>T$, the decay would take place in a matter-dominated era. However, as we will see below, the amount of energy imprinted on the ALP makes it incompatible with structure formation constraints.  To see that this is incompatible with fulfilling the correct DM abundance and the warmness constraint, let us note that the highest ratio between the DP and radiation energy densities happens just before the decay, around $\sim \Tdec$, such
\begin{align}
    \frac{\rho_{\gamma'}(\Tdec)}{\rho_\gamma(\Tdec)}=\xi\gg 1,
\end{align}
which leads to $
    \Tdec\sim {Y_\phi^\infty\,\mgamma}/{\xi}$. 
Now, the decay temperature should satisfy Eq.~\eqref{eq:tdec_warm}
which yields,
\begin{align}
    \frac{{Y_{\phi}^\infty}m_\phi}\xi>400\,\mbox{eV}.
\end{align}
Finally, replacing the above in the DM abundance, Eq.~\eqref{eq:DM_relic} gives
\begin{align}
    \Omega_{\phi,0}\gtrsim 500\,\xi^{1/4},
\end{align}
where we have taken the entropy dilution factor to be equal to $\xi^{-3/4}$. Since $\xi$ is equal or greater than 1, the ALP is always overproduced.

We now  analyze the situation where the cosmological evolution of our model could interfere with the process of BBN and the effective number of neutrinos. Leaving aside the decay of the DP for a moment, we will be in a scenario where the DP is heavier than 10 MeV, such that it is already non-relativistic when BBN starts. On the other hand, the ALP can be much lighter than that contributing to the relativistic energy density around BBN. This requires that the amount of energy is not in conflict with the effective number of neutrinos at that epoch. This condition is fulfilled  when the DM particles are cold before structure formation and account for the whole cold DM today.

However, the population from the decay could lead to several potential issues:
\begin{itemize}
    \item[ a)]If the decay of the DP happens before BBN, the produced ALPs are relativistic and could contribute to the effective number of neutrinos at that epoch. This applies for decay temperatures $\Tdec\gtrsim 1~$MeV. Neutrino physics develops in the standard way and we expect a possible increase in $N_{\rm eff}$.
    \item[b)] If the decay happens after neutrino decoupling (and far away from BBN process) the entropy injection into photons is not shared by neutrinos, lowering their relative temperature to photons. {This could happen for } $\Tdec< T_{\rm BBN}$, where $T_{\rm BBN}\sim 10$~keV is the temperature at which BBN has already set in. In this case, the lowering of the neutrino temperature imprints changes in their contribution to $N_{\rm eff}$ value at CMB decoupling, decreasing its value. 
    \item[c)] The most {intricate case occurs} when the decay happens {\it around} BBN at a temperature between 10 MeV and 0.01 MeV, because the decay affects the BBN process itself, by changing the baryon-to-photon ratio and the effective number of neutrinos, through the contribution from the relativistic ALP and a possible lowering of the neutrino temperature. 
\end{itemize} 
This complex study would require a dedicated analysis of the BBN process, incorporating the evolution of the abundance of the light elements. Such an analysis is beyond the scope of this paper. To restrict the interference during this period, we instead use the results from Ref.~\cite{Yeh:2024ors}. There, they examine the impact of a cosmic, matter-like species that is non-relativistic  during BBN, and decays around that time. They consider two different scenarios: the decays are either electromagnetic or dark. 
To construct a testable parameter they write the relative abundance of the dark particle ($X$), with respect to radiation as
\bea
\frac{\rho_{X}(T)}{\rho_r(T)}\Bigg|_{T=10\rm{MeV}}=\xi, 
\label{eq:ratio_olive}
\eea
where the temperature of 10 MeV sets an initial temperature before BBN begins. Their methodology aims to produce a global likelihood function 
by combining Big Bang Nucleosynthesis (BBN) likelihood functions (the BBN theory likelihood convolved with the observational likelihood) with the Cosmic Microwave Background (CMB) likelihood functions derived from Planck data. The resulting function depends on three independent input parameters: the baryon-to-photon ratio ($\eta$), the lifetime of the matter component ($\tau_X=\Gamma_X^{-1}$), and the density ratio of Eq.~\eqref{eq:ratio_olive}. The total likelihood can then be marginalized to yield single or joint likelihoods for any selected combination of these parameters.

%%%%%%%%
\begin{figure}[t!]
	\centering 
\includegraphics[scale=0.3]{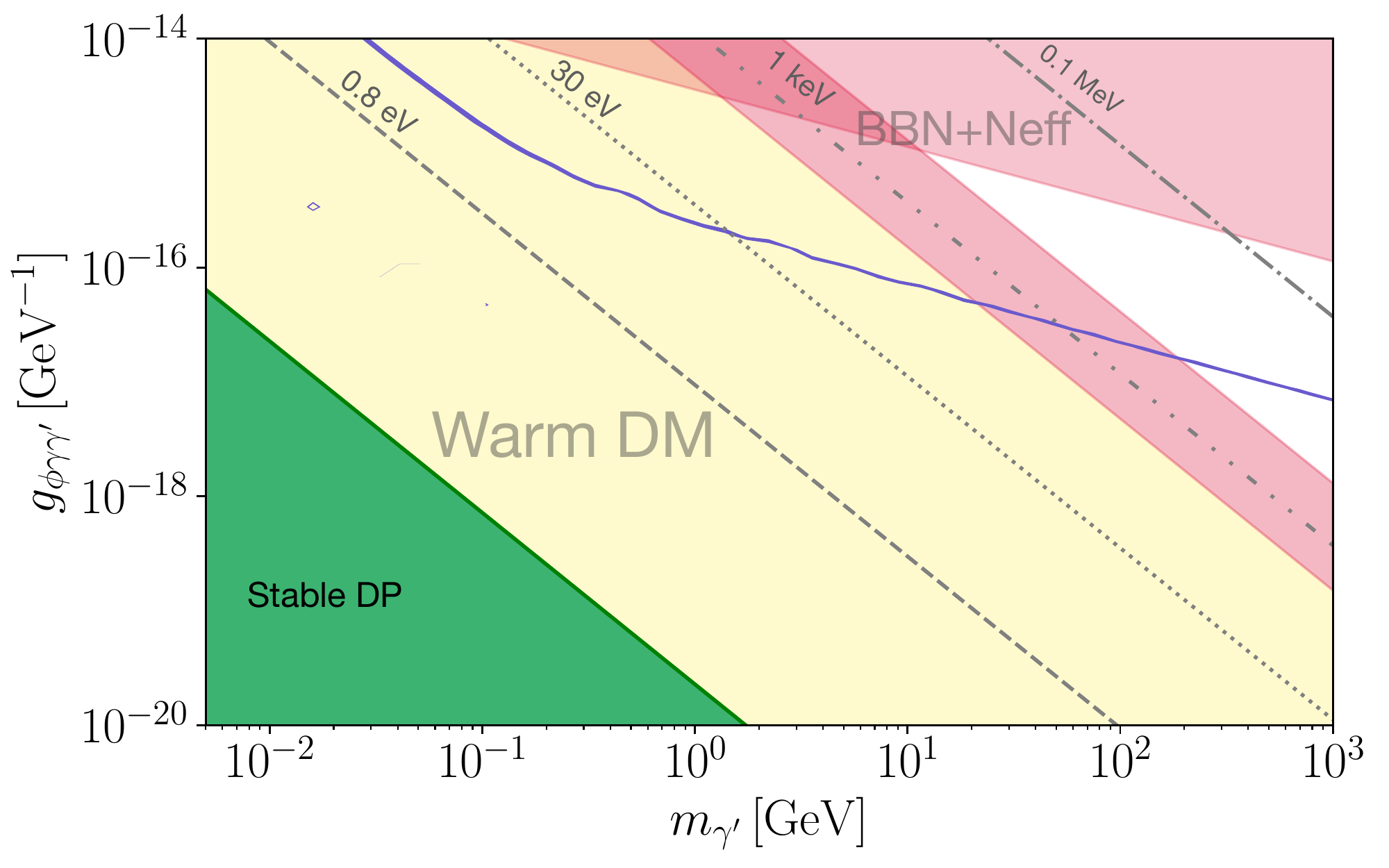}
\caption{Cosmological constraints on the parameter space coupling vs DP mass. We have assumed $m_\phi=10^{-2}\mgamma$. The segmented lines are isotherms for the decay temperature of the DP. The green area is where the DP is stable. The purple line indicates the values giving the measured DM abundance.} 
\label{fig:boundsofar}
 \end{figure}
 %%%%%%%%%%%%%%%

In the case of electromagnetic decay, their results show that there is a mild preference for non-zero perturbations, because of the slight preference in Ref.~\cite{Yeh:2024ors} for a value of $N_\nu$ smaller than $3$ in the CMB. The preferred $\xi$ value is around $\xi (\tau_X/1\,\mbox{ sec})^{1/2}\sim$\,0.015, for $\tau_X\sim 1\,\mbox{to}\,100$~sec. For dark decay, the constraints are mainly dominated by the value of $N_{\rm eff}$, affected by BBN data at small lifetimes and from CMB at higher ones (tightest constraints).

Our scenario is a mix between their two benchmark cases: on the one hand, the electromagnetic decays can decrease the effective number of neutrinos and the baryon-to-photon ratio, and increase the effective number of neutrinos through dark decay. Therefore, we take the following approach: we set a constant value of $\xi\lesssim 5\times 10^{-5}$ at $T_{\rm ini}=10$~MeV and we apply it for dark photon lifetimes smaller or equal than $10^4$ seconds, which corresponds to decays around $T=10$~keV, to avoid strong constraints on photodisintegration of light nuclei \cite{Cyburt:2002uv, Hufnagel:2017dgo, Hufnagel:2018bjp, Kawasaki:2020qxm, Depta:2020zbh}.

\subsection{{Results:} Standard Freeze-in parameter space}

A schematic plot showing the regions in the parameter space where the cosmological bounds constrain our model is presented in Fig.~\ref{fig:boundsofar}. In the green labeled ``Stable DP'' the DP is stable and therefore does not apply to the model presented here. The yellowish region ``Warm DM'' is the region where the ALP produced from the decay of the DP is too hot before matter radiation equality and therefore, would not properly account for structure formation. The pinkish ``BBN$+$Neff'' forbids the parameter space because decay would alter the process of BBN and/or give an effective number of neutrinos above the measured value. To produce this figure we have  fixed $\TRH=10^{10}$~GeV and $m_\phi=10^{-2} \mgamma$. The different lines show the decay temperature of the DP, for guidance.

%%%%%%%%
\begin{figure}[t!]
	\centering 
\includegraphics[scale=0.6]{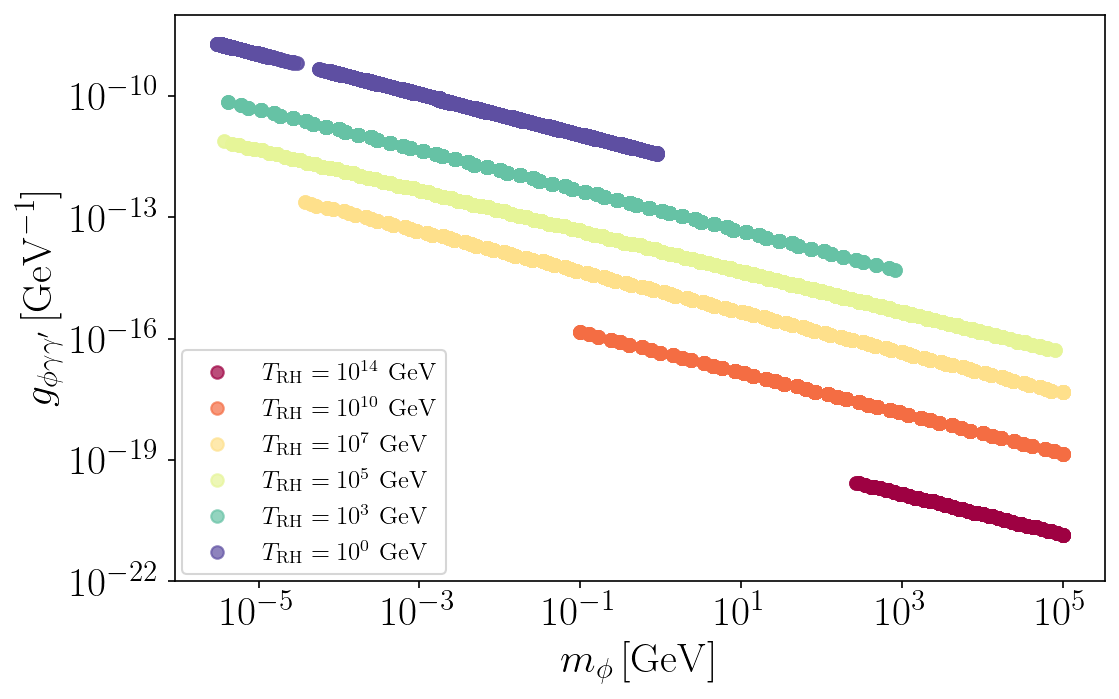}
\caption{Parameter space of the dark axion portal where the standard freeze-in gives ALP DM with the correct relic abundance and surviving all constraints mentioned in section \ref{sec:cosmo}. We have scanned over the dark photon mass, for several fixed reheating temperatures. We always respect the condition of standard reheating that $\TRH\geq \mgamma$. }
\label{fig:gvsmphi_TRH}
 \end{figure}

%%%%%%

In Fig.~\ref{fig:gvsmphi_TRH} we scan the parameter space $\Gg$ vs $\mphi$ for a set of reheating temperatures, with $\mgamma$ such that all the above-explained constraints are successfully satisfied. We have only scanned masses up to $\mphi\lesssim 10^5$~GeV and $\mgamma\lesssim 10^6$~GeV, respectively. The gap in the $\TRH=1$~GeV scan appears because the DP mass is restricted for that case to {be} $0.1\, \mbox{GeV}\lesssim\mgamma \lesssim 1\,\mbox{GeV}$, which for ALP masses ($\sim 60$~keV) fail to satisfy the required constraints.

%%%%%%%%
\begin{figure}[t!]
	\centering 
\includegraphics[scale=0.4]{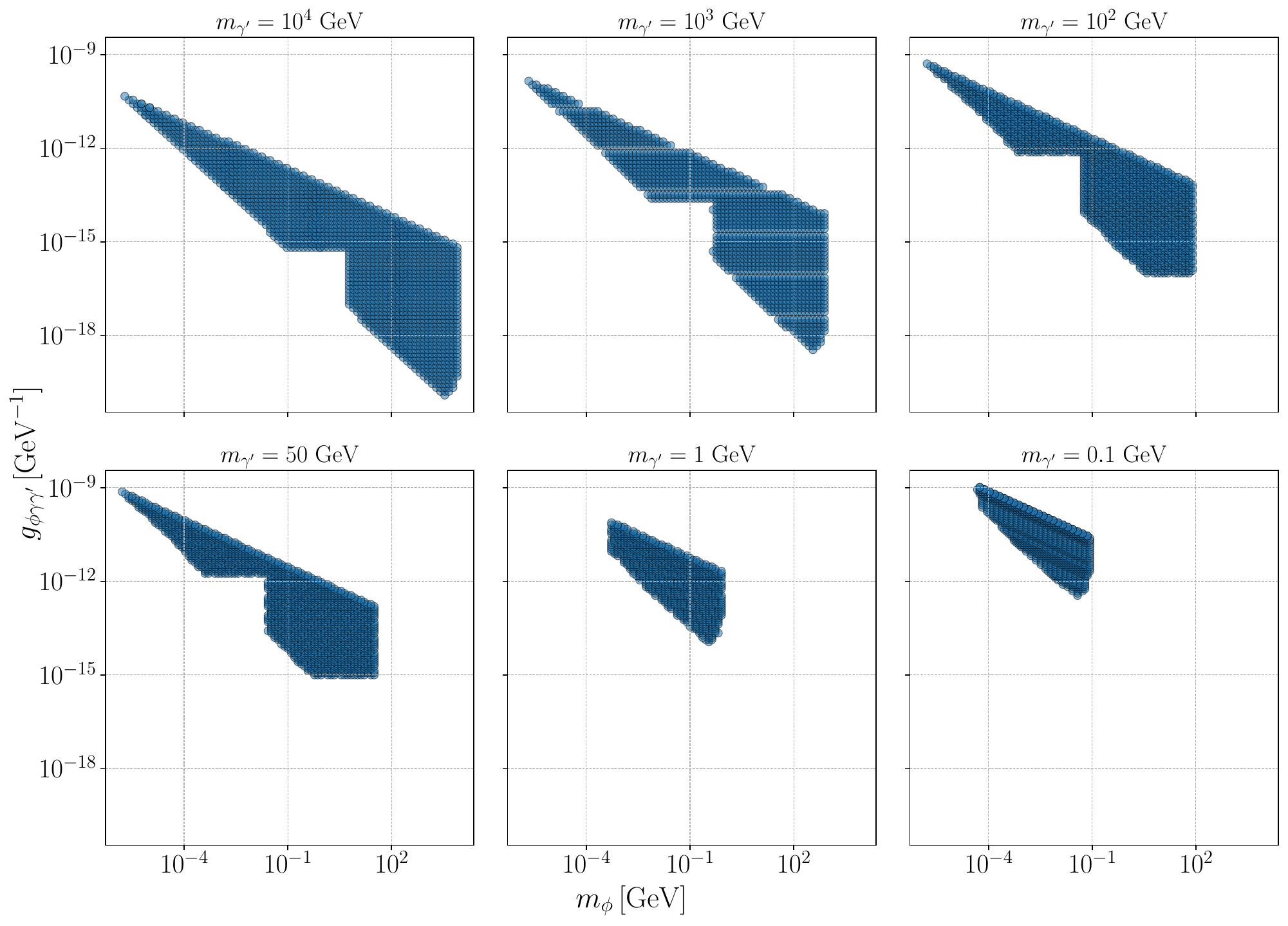}
\caption{Parameter space of the dark axion portal where the standard freeze-in gives ALP DM with the correct relic abundance and surviving all constraints mentioned in section 5. Each panel has a fixed dark photon mass (written above each plot) and we have blindly scanned over the reheating temperature $\TRH$, always respecting that {it} is above the masses of the dark particles.  }
\label{fig:gvsmphi_mgamma}
 \end{figure}

%%%%%%

On the other hand, in Fig.~\ref{fig:gvsmphi_mgamma} {show  the complementary plot, delineating } the parameter space of ($\Gg, \mphi$) that gives the correct relic abundance and has no conflict with cosmological observations for a set of dark photon masses, with a blind scan in $\TRH$ in the range of $10^{14} -0.1$~GeV, with the sole restriction that the scan is performed respecting $\TRH\geq\mgamma$. 

By inspecting figures \ref{fig:gvsmphi_TRH} and \ref{fig:gvsmphi_mgamma} we see that the range of masses where the ALP can satisfy the correct relic abundance is wide, {allowing} for different reheating temperatures. {High reheating scales only allow ALP dark matter with masses above the GeV scale and relatively low couplings $\Gg \lesssim \, 10^{-19}\,\mathrm{GeV}^{-1}$. This is because ALPs must become non-relativistic early enough to cool sufficiently before matter-radiation equality. Furthermore, high reheating temperatures increase the ALP yield, necessitating small couplings to achieve the correct relic abundance. Lower reheating temperatures broaden the ALP mass range to include smaller masses. }
The upper bound for the ALP mass is clearly  $\mphi\lesssim \TRH$. On the other hand, the smallest ALP mass can be deduced analytically. To do so, let us note that under the assumption that the entropy dilution is negligible (which is a good approximation as we have stated above), the relic abundance is independent of the DP mass. The requirement that the relic abundance satisfies the whole DM today, fixes the yield obtained at rehating as a function of the ALP mass. Putting in equations the above, we recall Eq.~\eqref{eq:DM_relic}, and we find
\begin{align}
{Y_{\phi}^\infty}=\frac{\Omega_{{cdm}}\, \rho_{c,0}}{s(T_0) \mphi},
\end{align}
with $\Omega_{ cdm}\sim 0.26$.
Using the expression \eqref{eq:initial_yield} for the yield, we find 
\begin{align}
\Gg^2=\frac{\Omega_{cdm}\,\rhoc}{s(T_0) \, \tilde Y\,\TRH\,\mphi},
\end{align}
with $\tilde Y\approx \frac{\textcolor{red}{1\times 10^{16}}}{\sqrt{\gs(\TRH)}\gss(\TRH)} \, $~GeV.
On the other hand, one of the strongest constraints is the warmness of the dark matter (see Fig.~\ref{fig:boundsofar}), which imposes that the decay temperature of the DP { satisfies} Eq.~\eqref{eq:tdec_warm}, which, we remind the reader, is given by
\begin{align}
   \Tdec\gtrsim \frac{T_{\rm eq}}{2
   \times 10^{-3}}\frac{\mgamma}{\mphi}, \,\,\,\,\,\,\,\mbox{with}\,\,\,\,\,\,\,\Tdec\simeq\frac{\Gg }{10}\sqrt{\frac{M_P\,\mgamma^3 }{\gs(\Tdec)^{1/2}}}.
\end{align}
Combining with the requirement of the relic abundance defines the following relation
\begin{align}
    \mphi^{1/2}\gtrsim 5 \times 10^3\,\frac{T_{\rm eq}\,\gs(\Tdec)^{1/4}}{\sqrt{\Omega_{cdm} \,\rho_{c,0}}} \sqrt{\frac{\tilde Y_{\rm RH}\, \TRH \,s(T_0)}{{M_P} \mgamma}}.
    \label{eq:min_mphi_1}
    \end{align}
 Replacing the rest of the parameters, we obtain
 \begin{align}
   \mphi\gtrsim 1.5 \times 10^{-6}\, \mbox{GeV} \,{\frac{\TRH}{\mgamma}} \left(\frac{\gs(\Tdec)}{\gs(\TRH)}\right)^{1/2} \left(\frac{106}{\gss(\TRH)}\right) .
\end{align}
Let us note that the highest possible DP mass in this standard reheating context is $\mgamma=\TRH$. In that case, the dependence on those variables cancels out in Eq. \eqref{eq:min_mphi_1} and the minimum ALP mass is almost insensible on the cosmology and the DP mass,  only mildly affected through the relativistic degrees of freedom.

\subsection{Strong freeze-in parameter space}
For the strong freeze-in case   the reheating temperature $\TRH$ is below both dark particle masses. To find the viable parameter space free of constraints, we make use of the yields for the different processes involved given in Eqs.~\eqref{analytic_1}-\eqref{analytic_3}. 

{In this scenario we expect the following evolution:} 
\begin{itemize}
\item Depending on the smallness of the ratio $\TRH/\mgamma$, the production of DPs is be exponentially suppressed, making their contribution to the early universe negligible. Numerically, we consider this to be the case when $Y_{\gamma'} < 10^{-20}$, with the yield defined in Eq.~\eqref{eq:gamma_yield}.
\item The decay of the DP takes place after a time $\tau=\Gamma_{\gamma'\rightarrow\phi\gamma}^{-1}$ {from reheating }. Depending on the coupling value $\Gg$ and the DP mass, this time can be almost immediately after reheating ({as} will be the case for high couplings) or not. 
\item If the decay temperature of the dark photon is smaller than $T_{\rm ini}=50$~MeV, we apply the BBN constraints.
\item We require the decay temperature of the DP to satisfy the warmness constraint, {\it i.e} it shall satisfy Eq.~\eqref{eq:tdec_warm}.
\item We always consider ALP masses higher than $\TRH$, therefore, we do not check their contribution to $\Delta N_{\rm eff}$.
\end{itemize}

%%%%%%%%%
\begin{figure}
\centering 
\includegraphics[scale=0.4]{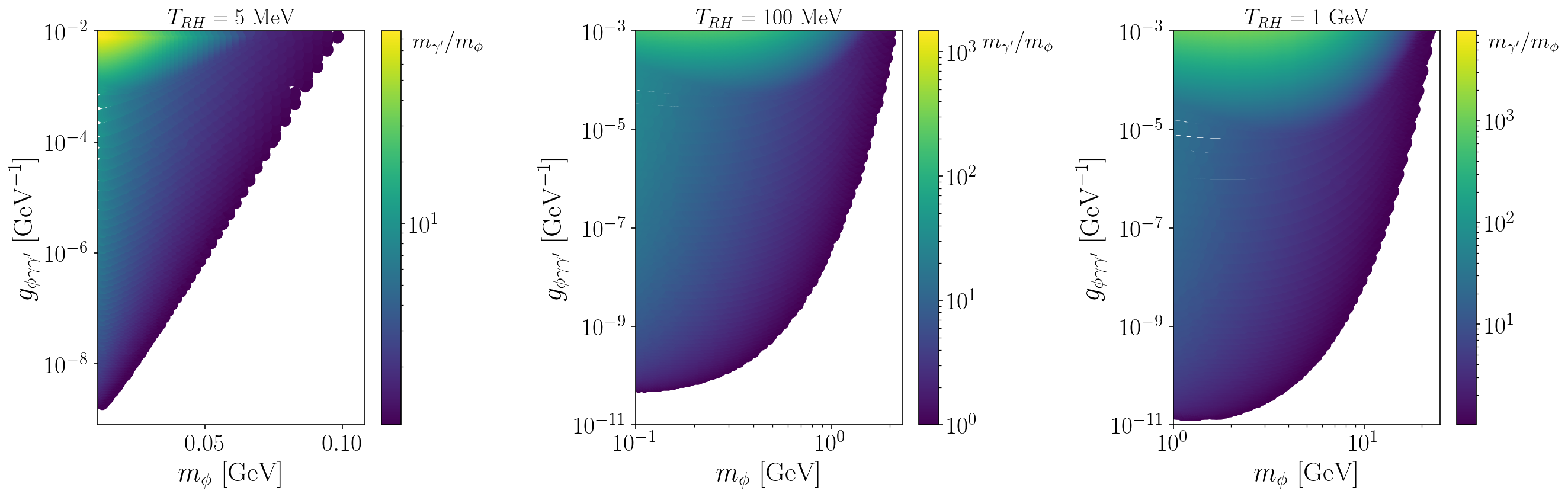}
\caption{Allowed parameter space for the strong freeze-in scenario in terms of the coupling $\Gg$ and the ALP mass $\mphi$ for different reheating temperatures.}
\label{fig:strongFI}
 \end{figure}
%%%%%%%

With the above in mind, we determine numerically the parameter space where the production of ALP dark matter is viable in the strong freeze-in scenario. In Fig.~\ref{fig:strongFI} we show our findings in the parameter space of the coupling $\Gg$ versus the ALP mass $\mphi$ for different reheating temperatures. 
When computing these plots, we assume that both dark masses are always greater than or equal to the corresponding reheating temperature. We can see that for similar ALP and DP masses not much higher than the reheating temperature, the allowed couplings are small and they increase as the hierarchy between the DP and the ALP increases. This can be understood as follows: When their masses have similar values and are relatively small (meaning close to $\TRH$), they are produced at the same rate, and production is dominant through annihilation of fermions. As the difference in mass increases, the DP production gets more suppressed, to the point that the fermion annihilation process is not efficient and the photon annihilation into two ALPs takes over. When the latter occurs, the process still depends on the DP mass (see Eq.~\eqref{analytic_1}) with higher masses suppressed by a power law.

%%%%%%%%%

\begin{figure}[t]
    \centering
    \begin{subfigure}[b]{0.48\textwidth}  % Adjust width as needed
        \centering
        \includegraphics[width=\textwidth]{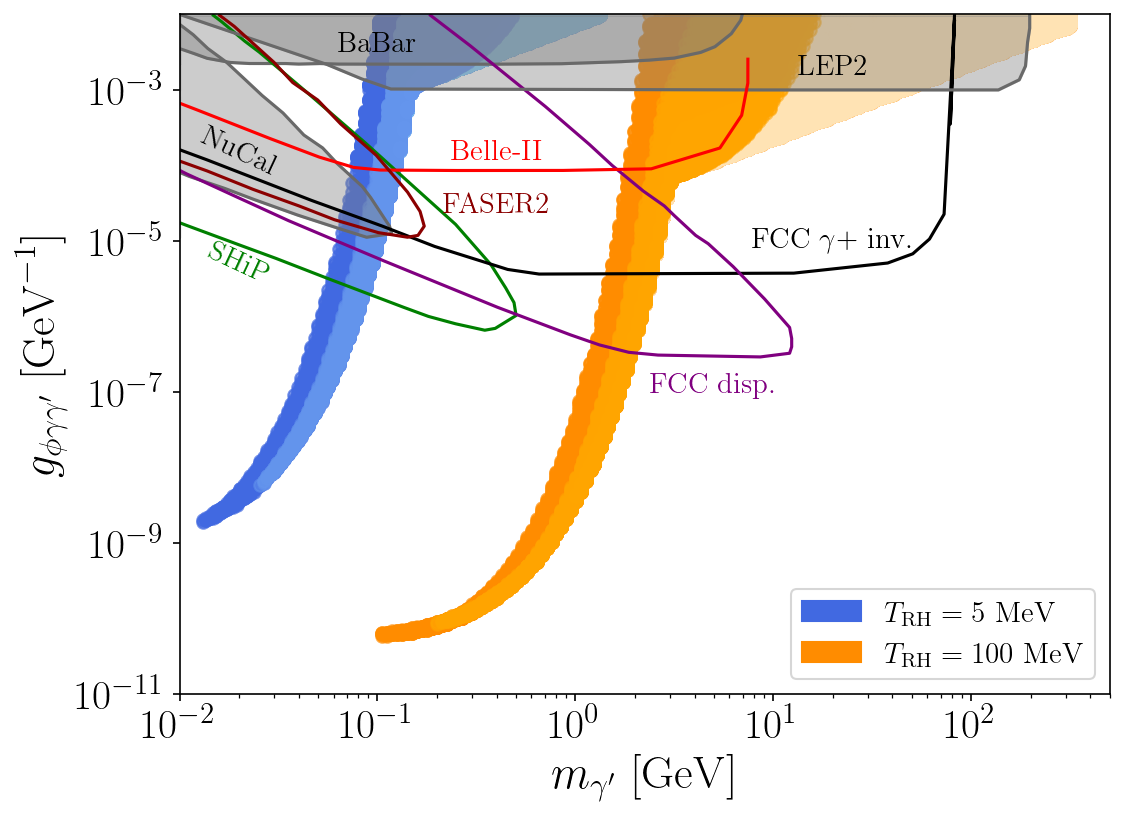}  
    \end{subfigure}
    \hfill
    \begin{subfigure}[b]{0.48\textwidth}  
        \centering
        \includegraphics[width=\textwidth]{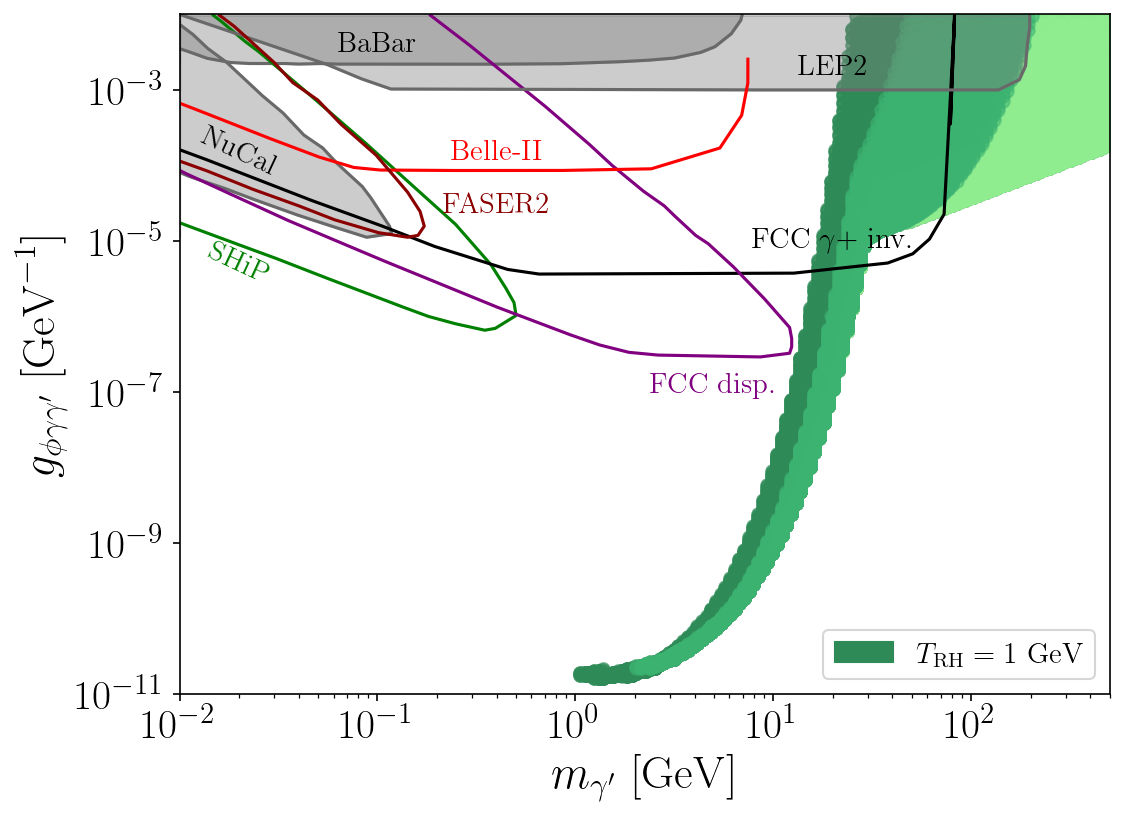}
        \end{subfigure}
        \caption{Allowed parameter space for the strong freeze-in scenario in terms of the coupling $\Gg$ and the DP mass $\mgamma$. The corresponding ALP mass for each point can be inferred from Fig.~\ref{fig:strongFI}. Different color regions correspond to different reheating temperatures. The light shade of blue, orange, and green corresponds to parameter space where $\mphi/\mgamma\leq 0.1$. The middle tone to $0.1<\mphi/\mgamma\leq 0.5$ and the darker one to $\mphi/\mgamma>0.5$. We have also included the upper limits to this model from B-factories, collider and beam dump experiments, taken from Refs.~\cite{deNiverville:2018hrc,Jodlowski:2023yne,Jodlowski:2024lab}. Left: Two lower values for the reheating temperature, $\TRH=5$~MeV and $\TRH=100$~MeV. Right: The same for a reheating temperature of $\TRH=1$~GeV.} 
    \label{fig:strongFI_collider}
\end{figure}

%%%%%%%%%%%%%%%%%%

Remarkably, the strong freeze-in scenario, can be tested in different laboratory experiments.
The sensitivity of B-factories, beam dump and collider experiments to the dark axion portal coupling was studied in  Ref.~\cite{deNiverville:2018hrc,Jodlowski:2023yne, Jodlowski:2024lab}, from where we adopted their results \footnote{We are very grateful to Krzysztof Jodłowski for pointing out to us more experiments in the reach of testing this model.}. Strictly speaking, they were obtained in the massless ALP limit. Considering the available phase space for decays we expect that they also apply to very good accuracy in the lighter blue, orange, and green regions of Fig.~\ref{fig:strongFI_collider}. In the intermediate and darker colored regions, we expect that some moderate and potentially more sizable deviations in the sensitivities may arise due to the finite mass of the ALP.  That said, there is considerable overlap between the dark matter regions and those testable in experiments.

\section{Conclusions}\label{sec:conclusions}
In this paper, we have studied the freeze-in production of axions and dark photons through a pure dark axion portal interaction $\Gg \phi F_{\mu\nu}\tilde F'^{\mu\nu}$, that can be motivated by a $Z_2$ coupling preventing other interactions such as kinetic mixing.\footnote{In principle, this still allows for a Higgs portal coupling, but we assume this to be small in the present paper.} We find that the lightest dark state (which we have chosen to be the ALP) can be a good dark matter candidate in large portions of the parameter space, depending on the reheating temperature. After remarking that a misalignment contribution can be made to be\footnote{For relatively high reheating temperatures in the standard freeze-in scenario this is indeed a choice and may require some tuning. There, it would be interesting to study a combination of both contributions similar to Ref.~\cite{Feruglio:2024dnc}.} quite small under relatively mild conditions, we have considered two possible freeze-in scenarios. Firstly, the standard FI mechanism, with the reheating temperature above both ALP and DP mass. The main results, taking into account also the cosmological constraints are shown in Figs.~\ref{fig:gvsmphi_TRH} and \ref{fig:gvsmphi_mgamma}. Second, strong FI, which can take place when the reheating temperature is below the dark masses, cf. Figs.~\ref{fig:strongFI} and \ref{fig:strongFI_collider}. In this case, production is suppressed as a result of the lack of energy in the thermal bath to produce dark particles. Notably, in this case, the coupling is sufficiently large that a significant region in parameter space can be tested independently in a laboratory-based search, e.g. at Belle-II, SHiP, and FASER2 (Fig.~\ref{fig:strongFI_collider} and Refs.~\cite{deNiverville:2018hrc,Jodlowski:2023yne,Jodlowski:2024lab}).

\section*{Acknowledgments}
We are grateful to Ariel Arza for his valuable contribution at early stages of this work. PA  acknowledges support from FONDECYT project 1221463 and 1251613. This article is based upon work from COST Action COSMIC WISPers CA21106, supported by COST (European Cooperation in Science and Technology). JJ also acknowledges support from the EU via ITN HIDDEN (No 860881). B.D.S has been funded by ANID Fondecyt postdoctorado 2022 N°3220566 and thanks the hospitality of Heidelberg University.

\appendix

\section{Unitarity}\label{app:unitarity}
The EFT description for momentum transfer close to the EW scale must respect the gauge symmetry of the SM, otherwise annihilation of gauge bosons into dark states, relevant for freeze-in, could give rise to processes growing with the energy \cite{Bell:2015sza}. In particular, the Lagrangian $\mathcal{L}_F$ in Eq.~\eqref{eq:portal} does not respect the full gauge symmetry. Therefore, the calculation of freeze-in at high reheating temperature must be carried out with $\mathcal{L}_B$. Let us comment in more detail on this fact.

Considering $\mathcal{L}_F$, the cross section for $W^+W^-\rightarrow \gamma'\phi$ is given by
\begin{eqnarray}\label{csww}
    \sigma(s) = \frac{e^2\Gg^2 s(12 m_W^2 + 20m_W^2 s + s^2)}{864\pi m_W^4 (s - 4m_W^2)}\left[\frac{(s - 4m_W^2)(m_\phi^4 + (s - m_{\gamma'}^2)^2 - 2m_{\phi}^2(s + m_{\gamma'}^2))}{s^3}\right]^{3/2}\,.
\end{eqnarray}
In the limit $s \gg 4m_W^2$, Eq.~\eqref{csww} reduces to $\sigma(s) \rightarrow \frac{e^2 g^2 s^2}{864\pi m_W^4}$, manifesting the growth in energy as $E_\text{CM}^4$.

Let us turn to the EFT that is $SU(2)\times U(1)$ symmetric, $\mathcal{L}_{B}$. Now, the annihilation process $W^+W^- \rightarrow \phi \gamma'$ is mediated by both the SM photon and the $Z$ boson (or equivalently by the $B_\mu$ boson from $U(1)_Y$). The cross section in this case is given by,
\begin{eqnarray}
    \sigma(s) = \frac{c_w^2e^2\Gg^2 m_Z^4s(12m_W^4 + 20m_W^2 s + s^2)}{864m_W^4\pi (s- m_Z^2)^2(s - 4m_W^2)}\left(\frac{s - 4m_W^2}{s} \right)^{3/2}\xrightarrow[]{s\rightarrow \infty} \text{constant}\,,
\end{eqnarray}
and the problematic fast growth disappears.\footnote{We note that it does not decay as $\sim 1/s$. This is due to the nonrenormalizable coupling $\Gg$.}
In this way, using the symmetric gauge EFT description $\mathcal{L}_B$, we can calculate the freeze-in abundance of relics up to high values of $T_{RH}\lesssim \Gg^{-1}$.

\section{Cross sections}
The differential cross section of the conversion process in the centre of mass frame is given by
\begin{align}\label{master}
    \left(\frac{d\sigma_{12\to 34} v}{d\Omega}\right)_{c.o.m.} = \frac{|\mathbf{p_f}|}{64\pi^2 E_1 E_2 \sqrt{s}} \braket{|\mathcal{M}_{12\to 34}|^2},
\end{align}
where $\sqrt{s}$ denotes the total c.o.m. energy, $|\mathbf{p_f}|=\lambda(s,m_3^2,m_4^2)^{1/2}/(2\sqrt{s})$ denotes the final state momentum\footnote{$\lambda(x,y,z)=(x-y-z)^2-4yz$ is the Källén function.} and
$E_1=(|\mathbf{p_i}|^2+m_1^2)^{1/2}$ and $E_2=(|\mathbf{p_i}|^2+m_2^2)^{1/2}$ denote the energies of the initial state DM and SM particle 
with momentum $|\mathbf{p_i}|= \lambda(s,m_1^2,m_2^2)^{1/2}/(2\sqrt{s})$. 

In the following, we make use of expression \eqref{master} to calculate some cross sections times relative velocity for relevant processes involved in the freeze-in analytic calculation. Moreover, in some parts we also make use of the non-relativistic annihilation of dark states to expand in the relative velocity and then retaining the $s$ and/or $p$-wave, making use of \calchep\ code \cite{Belyaev:2012qa}.

\subsection{$f\bar{f}\rightarrow \phi\gamma'$}\label{app_A1}
Let us start with the charged lepton annihilation, and the rest of fermions follow in a similar way. Defining the momenta as $l(p_1) + \bar{l}(p_2) \rightarrow \phi(p_3) + \gamma'(p_4)$, with $l = e,\mu,\tau$, we have
\begin{align}
    t = (p_1-p_3)^2 = m_e^2 + m_\phi^2 - 2 (|\mathbf{p_i}|^2+m_e^2)^{1/2} (|\mathbf{p_f}|^2+m_\phi^2)^{1/2} + 2 |\mathbf{p_f}||\mathbf{p_i}| \cos\theta.
\end{align}
Besides, in the ultra-relativistic regime, $\sqrt{s}\gg (m_l, m_{\phi}, m_{\gamma'})$, we obtain that the matrix element reduces to
\begin{eqnarray}\label{matele_1}
    \braket{|\mathcal{M}_{f\bar{f}\rightarrow \phi\gamma'}|^2} = e^2\Gg^2\left(\frac{s^2 + 2st + 2t^2}{s}\right)\,.
\end{eqnarray}
Furthermore, in this case we have $t = \frac{s}{2}(-1 + \cos(\theta))$. Plugging Eq.~\eqref{matele_1} into Eq.~\eqref{master}, we obtain
\begin{eqnarray}
    \sigma v(s) &=& \frac{e^2\Gg^2}{96m_W^2m_Z^2\pi(s - m_Z^2)^2}[8m_W^4(s - m_Z^2)^2 + 4(-3 + 4c_W^2)m_W^2m_Z^2s(s - m_Z^2) \\ &&\qquad\qquad\qquad\qquad\qquad\qquad\qquad\qquad\qquad\qquad\qquad+ (5 - 12c_W^2 + 8c_W^4)m_Z^4 s^2] \nonumber
\end{eqnarray}
where $c_W$ is the cosine of the Weinberg angle. At very high energies, $\sigma v$ for the charged leptons becomes a constant, such that the 
\begin{eqnarray}
\braket{\sigma v}_{l\bar{l}\rightarrow
    \phi\gamma'} = \frac{5e^2\Gg^2}{96\pi c_W^2}\,.
\end{eqnarray}
Similarly, for up-type and down-type quarks we obtain, 
\begin{eqnarray}
    \braket{\sigma v}_{u\bar{u}\rightarrow
    \phi\gamma'} &=& \frac{17e^2\Gg^2}{2592\pi c_W^2}, \\
    \braket{\sigma v}_{d\bar{d}\rightarrow
    \phi\gamma'} &=& \frac{5e^2\Gg^2}{2592\pi c_W^2}\,.
\end{eqnarray}
However, for SM neutrinos, the $s$-channel is mediated only by the $Z$ boson, since they do not couple to the photon. At high energies we find,
\begin{eqnarray}
    \braket{\sigma v}_{\nu\bar{\nu}\rightarrow
    \phi\gamma'} = \frac{e^2\Gg^2}{24\pi c_W^2}\,.
\end{eqnarray}

\subsection{$\phi\gamma' \rightarrow f\bar{f}$}\label{app_strongfi1}
In the following, we compute the average annihilation cross section times velocity for $\phi\gamma' \rightarrow f\bar{f}$, with $f$ any SM fermion, assuming that the coannihilation is non-relativistic. This result is useful for the calculation of the relic abundance in the strong freeze-in in Sec.~\ref{strong_fi} considering $\mathcal{L}_B$ EFT. In each case, we have replaced $\braket{v^2}  \rightarrow 6T/\mu$, with $\mu = m_\phi m_{\gamma'}/(m_\phi  + m_{\gamma'})$ \cite{Toma:2013bka}. 

In the case of charged leptons in the final state and neglecting the fermion mass, 
\begin{eqnarray}
    \braket{\sigma v}_{\phi\gamma' \rightarrow l\bar{l}} = \frac{e^2 \Gg^2 \left[5 (m_\phi + m_{\gamma'})^4 - 12c_W^2(m_\phi + m_{\gamma'})^2m_Z^2 + 8c_W^4 m_Z^4\right]}{24 \pi c_W^2 \sqrt{m_\phi m_{\gamma'}}\left[(m_\phi + m_{\gamma'})^2 - m_Z^2\right]^2} T\,.
\end{eqnarray}
In the case of SM neutrinos, the mediator is only the $Z$ boson, and we obtain
\begin{eqnarray}
    \braket{\sigma v}_{\phi\gamma' \rightarrow \nu\bar{\nu}} = \frac{ e^2 \Gg^2 (m_\phi + m_{\gamma'})^4 }{24 \pi c_W^2 \sqrt{m_\phi m_{\gamma'}} ((m_\phi + m_{\gamma'})^2 - m_Z^2)^2} T
\end{eqnarray}
For up-type quarks we have
\begin{eqnarray}
    \braket{\sigma v}_{\phi\gamma' \rightarrow u\bar{u}} = \frac{e^2 \Gg^2 \left[17 (m_\phi + m_{\gamma'})^4 - 40c_W^2(m_\phi + m_{\gamma'})^2 m_Z^2 + 32c_W^4 m_Z^4\right]}{72 \pi c_W^2 \sqrt{m_\phi m_{\gamma'}} \left[(m_\phi + m_{\gamma'})^2 - m_Z^2\right]^2} T,
\end{eqnarray}
and for down-type quarks
\begin{eqnarray}
    \braket{\sigma v}_{\phi\gamma' \rightarrow d\bar{d}} = \frac{e^2 \Gg^2 \left[5 (m_\phi + m_{\gamma'})^4 - 4c_W^2(m_\phi + m_{\gamma'})^2 m_Z^2 + 8c_W^4   m_Z^4\right]}{72 \pi c_W^2 \sqrt{m_\phi m_{\gamma'}}\left[(m_\phi + m_{\gamma'})^2 - m_Z^2\right]^2} T.
\end{eqnarray}

\subsection{$W^+W^-\rightarrow \phi\gamma'$}
In the following, we calculate the average annihilation cross section times velocity for the $W$ boson annihilation in the non-relativistic limit, $\sigma v$ for $\sqrt{s} \lesssim m_W$. Defining the momenta as $W^-(p_1) + W^+(p_2) \rightarrow \phi(p_3) + \gamma'(p_4)$ we have that,
\begin{align}
    t = (p_1-p_3)^2 = m_W^2 + m_\phi^2 - 2 (|\mathbf{p_i}|^2+m_W^2)^{1/2} (|\mathbf{p_f}|^2+m_\phi^2)^{1/2} + 2 |\mathbf{p_f}||\mathbf{p_i}| \cos\theta.
\end{align}
After some algebra and integrating over $t$ we obtain,
\begin{eqnarray}
    \sigma v(s) = \frac{e^2\Gg^2}{432\pi m_W^4 s}\left[3s^3 -48m_W^6 - 2s^3 + 16m_W^2 s^2 - 68m_W^4 s \right]\,.
\end{eqnarray}
Applying the non-relativistic approximation to obtain the average annihilation cross section, $\braket{\sigma v} = \sigma v(s = \braket{s})$, with $\sqrt{s} = 4m_W^2(1 + \braket{v^2}/4)$, we obtain an expansion in velocities $\braket{\sigma_{WW\rightarrow\phi\gamma'} v} = \frac{e^2 \Gg^2}{16\pi} v^2 + \mathcal{O}(v^4)$. In terms of temperature, we have that for a Maxwell-Boltzmann distribution, $\braket{v^2} \rightarrow 6T/m_W$, such that 
\begin{eqnarray}  
\braket{\sigma_{WW\rightarrow\phi\gamma'} v} = \frac{3e^2 \Gg^2}{8\pi}\frac{T}{m_W} + \mathcal{O}(T^2)\,.
\end{eqnarray}

\bibliography{bibliography}
\bibliographystyle{utphys}
\end{document}